\documentclass[11pt]{article}

\usepackage[a4paper,margin=2.35cm]{geometry}
\usepackage[T1]{fontenc}
\usepackage[utf8]{inputenc}
\usepackage{microtype}
\usepackage{amsmath}
\usepackage{booktabs}
\usepackage{siunitx}
\usepackage{xcolor}
\usepackage{graphicx}
\usepackage{caption}
\usepackage{placeins}
\usepackage{setspace}
\usepackage[hidelinks]{hyperref}
\usepackage{url}

\title{\textbf{A finite-support convex Massieu approximation for global phase equilibrium}}
\author{
Piyang Liu$^{1}$, Kai Zhang$^{1,*}$, Shuyu Sun$^{2,\dagger}$, Xia Yan$^{3}$ and Jun Yao$^{3}$\\
\small $^{1}$School of Artificial Intelligence, Qingdao University of Technology, Qingdao, China\\
\small $^{2}$School of Mathematical Sciences, Tongji University, Shanghai, China\\
\small $^{3}$School of Petroleum Engineering, China University of Petroleum (East China), Qingdao,
China\\
\small $^{*}$Corresponding author: Kai Zhang,
\href{mailto:zhangkai@qut.edu.cn}{zhangkai@qut.edu.cn};\\
\small $^{\dagger}$Corresponding author: Shuyu Sun,
\href{mailto:suns@tongji.edu.cn}{suns@tongji.edu.cn}
}
\date{}

\begin{document}
\maketitle

\begin{abstract}
Compositional multiphase flow simulations typically recover phase equilibrium through local
flash calculations repeated for successive thermodynamic states. Here we show that phase
stability, phase partitioning and thermodynamic properties can instead be organized around
phase-specific convex Massieu potentials constructed from finite sets of native thermodynamic
support states. Each potential yields pressure through its value and generates internal-energy
and component molar densities through its gradient, while competition between potentials
determines phase stability and coexistence. Benchmarks spanning vapour--liquid and
liquid--liquid equilibria, isothermal and nonisothermal conditions, and cubic and
multiparameter equations of state and a composite gas--brine model show agreement with
independent reference calculations. In fully implicit ternary flow, the approximation
captures repeated phase appearance and disappearance without nested local flashes, reproducing
the reference phase state in every cell at every accepted time step. These results establish
offline-constructed convex potentials as a reusable thermodynamic closure within prescribed
state domains.
\end{abstract}

Compositional transport models conventionally combine an equation of state (EOS), which
relates pressure, composition and phase properties\cite{coats1980,nghiem1981,peng1976},
with local stability and phase-split calculations that identify the stable phases and
partition the components among them\cite{michelsen1982a,michelsen1982b,michelsen2007}.
Flash formulations with alternative state-function and volume specifications use different
thermodynamic coordinates\cite{michelsen1999,mikyska2011,smejkal2017}, yet share a common
computational task: recovering equilibrium at each queried state. In conventional flash-based
flow simulations, this task is repeated across grid cells, nonlinear iterations and time
steps as the conservation equations are solved.

Existing approaches address this repeated computation at different levels. Accelerated
successive substitution\cite{wang2022gdem} reduces the computational work of individual
flash solves. Reuse of equilibrium information offers a complementary route: compositional-space
and tie-simplex parameterizations exploit the geometry of equilibrium
states\cite{voskov2008,voskov2009,iranshahr2010}, and sparse-grid surrogates approximate
equilibrium from precomputed data\cite{wu2015}. Operator-based linearization extends reuse
to state-dependent operators\cite{khait2017}, and learned approximations have been applied
to phase coexistence and compositional flow\cite{dhamankar2025,fang2026}. Fully implicit
VT-flash formulations\cite{li2022}, persistent unknowns\cite{lipovac2024} and global
complementarity conditions\cite{lauser2011} reorganize how equilibrium is coupled to
transport. Together, these approaches bring two related aspects of thermodynamic closure
into focus: how equilibrium information is represented and how it is integrated into the
conservation problem. On the representation side, thermodynamically consistent learning
of excess Gibbs energy illustrates a generating principle: a single scalar potential
generates thermodynamically consistent activity coefficients through
differentiation\cite{hoffmann2026}.

The convex geometry of equilibrium provides a basis for extending this generating principle
to phase-specific potentials. At fixed temperature and pressure, tangent-plane stability
analysis tests whether the tangent plane at a homogeneous state supports the Gibbs-energy
surface from below; Gibbs-energy minimization identifies stable combinations of phase
states; and common-tangent and convex-envelope constructions reveal the corresponding
supporting states. Convex-hull formulations make this geometry explicit and recover the
number and compositions of stable phases\cite{hildebrandt1994,gottl2023,gottl2025}. The
same geometry admits a dual description through thermodynamic dual extremum
principles\cite{mitsos2007} and dual optimization in volume--composition space\cite{pereira2010},
and a thermodynamically stable flash solution recovers the corresponding equilibrium through
equality of intensive variables and material balance. Repeated local calculations therefore
recover different parts of a common equilibrium geometry, one queried state at a time.
This motivates a direct, structure-preserving representation of that geometry that can be
evaluated and reused.

Such a representation requires coordinates that are shared by competing phases and compatible
with conservation equations. Temperature and component chemical potentials, which are equal
in coexisting phases, provide such coordinates, and free-energy and chemical-potential
formulations supply a thermodynamic basis for multicomponent and nonisothermal
transport\cite{kou2018cma,kou2018entropy}. The entropy representation gives the required
conjugate variables: for a state $\boldsymbol q$ comprising internal-energy and component
molar densities, the reduced negative entropy density $h(\boldsymbol q)=-s(\boldsymbol
q)/R$ is convex over stable equilibrium states, and its conjugate coordinates are
$\eta_0=-1/(RT)$ and $\eta_i=\mu_i/(RT)$\cite{callen1985}. The Legendre--Fenchel
conjugate\cite{rockafellar1970} defines a Massieu potential that equals the reduced negative
grand-potential density, $p/(RT)$\cite{alberty2001}. Phase potentials can thus be compared
at shared dual coordinates, and the Massieu potential becomes a single object linking
equilibrium geometry, state generation and conservation closure.

Here we introduce the \emph{Massieu Approximation for Global Phase Equilibrium} (MAGE),
which realizes this dual geometry through finite collections of native supporting planes.
Each phase is represented by a differentiable convex potential that generates pressure
through its value and thermodynamic densities through its gradient, and competition between
potentials determines stability and coexistence among the represented phases. The finite
construction preserves convexity and the thermodynamic generating identities, and conditional
error bounds connect support coverage, smoothing and phase-potential competition to
equilibrium accuracy. For a fixed reference model and prescribed state domain, the potentials
are constructed offline and coupled to the conservation equations through phase complementarity,
providing reusable thermodynamic closure. We assess the accuracy and compactness of the
representation, test its applicability across temperatures, reference models and phase
types, and demonstrate direct closure in fully implicit ternary flow with repeated phase
appearance and disappearance.

\section*{Results}

\subsection*{Finite-support Massieu representation}

To represent equilibrium geometry directly, we construct phase-specific potentials in
shared dual coordinates. The internal-energy density $u$ (J\,m$^{-3}$) and component molar
densities $c_i$ (mol\,m$^{-3}$) form the state vector
\begin{equation}
\boldsymbol q=(u,c_1,\ldots,c_{N_{\mathrm c}})^{\mathsf T}.
\label{eq:qdef}
\end{equation}
The Massieu coordinates are
\begin{equation}
\eta_0=-\frac{1}{R T},\qquad
\eta_i=\frac{\mu_i}{R T}\quad(i=1,\ldots,N_{\mathrm c}),
\label{eq:eta}
\end{equation}
where $R$ is the universal gas constant, $T$ is absolute temperature and $\mu_i$ is the
chemical potential of component $i$. For a differentiable homogeneous phase $\alpha$
satisfying the contact conditions below, the exact Massieu potential satisfies
\begin{equation}
\Phi_\alpha(\boldsymbol\eta)=\frac{p_\alpha}{R T},
\qquad
\nabla_{\boldsymbol\eta}\Phi_\alpha=\boldsymbol q_\alpha,
\label{eq:massieu_generator}
\end{equation}
generating pressure and the density state (Methods and Supplementary Note~1).

Each native reference state in a represented homogeneous phase domain defines an affine
lower bound on the restricted conjugate, which becomes a contact tangent when the state
attains the supremum (Methods and Supplementary Note~1). A support state $m$ on branch
$\alpha$ stores
$(\boldsymbol\eta_{\alpha m},\Phi_{\alpha m},\boldsymbol q_{\alpha m})$ and defines the plane
\begin{equation}
L_{\alpha m}(\boldsymbol\eta)=
\Phi_{\alpha m}+\boldsymbol q_{\alpha m}^{\mathsf T}(\boldsymbol\eta-\boldsymbol\eta_{\alpha m}).
\label{eq:support}
\end{equation}
To smooth the convex, generally nonsmooth maximum of $M$ planes, MAGE uses a normalized
log-sum-exp\cite{nesterov2005,calafiore2020},
\begin{equation}
\widehat\Phi_\alpha(\boldsymbol\eta)=\Delta\Phi+
\tau_\alpha\log\left[
\frac{1}{M_\alpha}\sum_{m=1}^{M_\alpha}
\exp\left(\frac{L_{\alpha m}(\boldsymbol\eta)}{\tau_\alpha}\right)
\right],
\label{eq:mage}
\end{equation}
where $\Delta\Phi$ is a common additive calibration. Within each branch (suppressing the
subscript on $\tau$), the normalized weights are
\begin{equation}
\pi_{\alpha m}=
\frac{\exp(L_{\alpha m}/\tau)}{\sum_j\exp(L_{\alpha j}/\tau)}.
\label{eq:pi}
\end{equation}
Differentiation gives the gradient as the weighted mean of the support states,
\begin{equation}
\nabla_{\boldsymbol\eta}\widehat\Phi_\alpha
=\sum_m\pi_{\alpha m}\boldsymbol q_{\alpha m},
\label{eq:gradmage}
\end{equation}
and its positive semidefinite Hessian is a scaled weighted covariance:
\begin{equation}
\nabla_{\boldsymbol\eta}^2\widehat\Phi_\alpha
=\frac{1}{\tau}\operatorname{Cov}_{\pi_\alpha}(\boldsymbol q_{\alpha m})\succeq0.
\label{eq:hessmage}
\end{equation}

Since $\pi_{\alpha m}\geq0$ and $\sum_m\pi_{\alpha m}=1$, every generated density state
lies in the convex hull of the native support states. In particular, nonnegative support
concentrations yield nonnegative generated concentrations. The generating identities are
preserved by construction: at fixed temperature $\partial\widehat p_\alpha/\partial\mu_i=\widehat
c_i^\alpha$, and the symmetry of the Hessian gives $\partial\widehat
c_i^\alpha/\partial\eta_j=\partial\widehat c_j^\alpha/\partial\eta_i$ and $\partial\widehat
c_i^\alpha/\partial\eta_0=\partial\widehat u_\alpha/\partial\eta_i$.

\begin{figure}[!htbp]
\centering
\includegraphics[width=\linewidth]{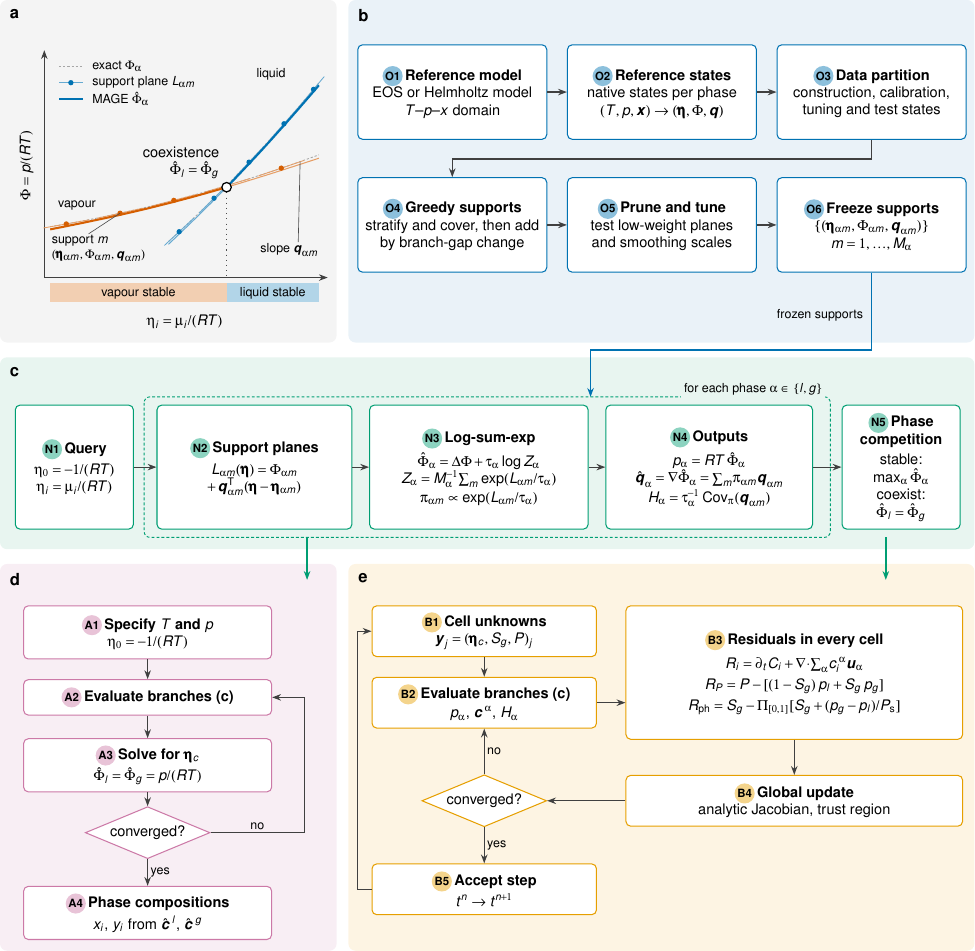}
\caption{\textbf{Construction and online use of finite-support Massieu potentials.}
\textbf{a} Stored support states define supporting planes
whose normalized log-sum-exp forms a smooth, convex phase branch. The larger potential
identifies the stable branch, and branch intersections define coexistence. \textbf{b}
Native tangent planes are
selected by stratified coverage and the change in the potential difference between competing phases.
Pruning and smoothing are evaluated on thermodynamic calibration and tuning states.
\textbf{c} Online evaluation returns the
branch potential, gradient and Hessian directly from the frozen supports. \textbf{d} Stand-alone
binary coexistence calculation at specified $T$ and $p$. \textbf{e} Fully coupled flow calculation,
in which the component balances, pressure closure and phase complementarity are solved using
analytic branch derivatives.}
\label{fig:concept}
\end{figure}

Distinct phases are represented by Massieu branches on a common dual domain. At fixed
$\boldsymbol\eta$, the largest potential identifies the stable phase among those represented:
\begin{equation}
\widehat\Phi_{\mathrm{eq}}(\boldsymbol\eta)=\max_\alpha\widehat\Phi_\alpha(\boldsymbol\eta).
\label{eq:phasemax}
\end{equation}
Active phases $\alpha$ and $\beta$ coexist where
$\widehat\Phi_\alpha=\widehat\Phi_\beta=\widehat\Phi_{\rm eq}$ at a common $\boldsymbol\eta$.
The shared coordinates impose equal temperature and chemical potentials, and
equality of the branch potentials imposes equal pressure. Although each phase branch is
differentiable and convex, their pointwise maximum is generally nondifferentiable at coexistence.
Through convex conjugacy, the mixture density is characterized as a convex combination of
the active branch gradients. At prescribed temperature, pressure and overall mole fractions,
the same closure follows from maximizing the dual objective $\boldsymbol\eta_c^{\mathsf
T}\boldsymbol z$ subject to $\widehat\Phi_\alpha\le P/(RT)$ for all represented phases;
the phase volumes per total mole are the associated nonnegative multipliers. For exact
branches, the dual objective bounds the reduced molar Gibbs energy from below (Supplementary
Note~1). Material balance and phase complementarity thus follow from the optimality
conditions of a common convex representation and enter the global conservation equations
directly in the coupled formulation. Figure~\ref{fig:concept} summarizes the offline
construction of the phase branches and their online use in stand-alone equilibrium and
coupled conservation calculations.

The finite-support construction also admits deterministic error bounds, stated here for
the uncalibrated potentials. The common calibration preserves gradients and phase differences
but contributes to absolute potential and pressure errors. The gap between the exact branch
and each contact support plane is its Bregman divergence\cite{bregman1967}. In the
dimensionless scaled coordinates used for support
construction (Methods), let $\varphi(\boldsymbol X)$ denote the exact branch, let $\Lambda$ bound
the norm of its Hessian and let the supports form a $\delta$-cover of the represented domain. The
potential error of the hard tangent envelope is then at most $\Lambda\delta^2/2$, and smoothing by
the normalized log-sum-exp adds at most $\bar\tau\log M$, where $\bar\tau=\tau/\Phi_s$ is the
smoothing scale of the dimensionless potential (Supplementary Note~2). If the branch is,
in addition, $\sigma$-strongly convex on a region containing the segments from the query
to all retained supports, the error in the scaled gradient that defines the generated state
satisfies
\begin{equation}
\left\|\nabla\widehat\varphi-\nabla\varphi\right\|
\le \Lambda\sqrt{\frac{\Lambda\delta^2+2\bar\tau\log M}{\sigma}}.
\label{eq:main_error_bound}
\end{equation}

For two competing branches $l$ and $g$ in a common scaled coordinate system and potential
reference, let $e_\alpha=\varphi_\alpha-\widehat\varphi_\alpha^{(0)}$ denote the uncalibrated
potential deficit, with $0\le e_\alpha\le\varepsilon_\alpha$ for $\alpha\in\{l,g\}$. Define
the exact and approximate phase margins as $\Delta=\varphi_l-\varphi_g$ and
$\widehat\Delta=\widehat\varphi_l-\widehat\varphi_g$. The common calibration cancels from
their difference, giving
\begin{equation}
\widehat\Delta-\Delta=e_g-e_l,\qquad
|\widehat\Delta-\Delta|\le\max\{\varepsilon_l,\varepsilon_g\}.
\label{eq:main_margin_bound}
\end{equation}
Stable-phase identification is preserved at shared dual coordinates wherever
$|\Delta|>\max\{\varepsilon_l,\varepsilon_g\}$. Along a transverse path, if the margin
slope has magnitude at least $\kappa>0$ between the exact and approximate crossings, their
separation in the path parameter is at most $\max\{\varepsilon_l,\varepsilon_g\}/\kappa$.
At prescribed pressure, the conditioning of the equilibrium equations governs the displacement
of the coexistence coordinates, whereas endpoint separation governs phase-fraction sensitivity
(Supplementary Note~2). Near critical points, coalescing phases, potentially increased
curvature and vanishing $\kappa$ weaken these bounds, motivating finer coverage. We therefore
test branch values, generated density states and phase-potential differences as complementary
measures of accuracy; the liquid--liquid benchmark also tests the potential-error bound directly.

\subsection*{Accuracy and support count in binary vapour--liquid equilibrium}

\begin{figure}[!htbp]
\centering
\includegraphics[width=\linewidth]{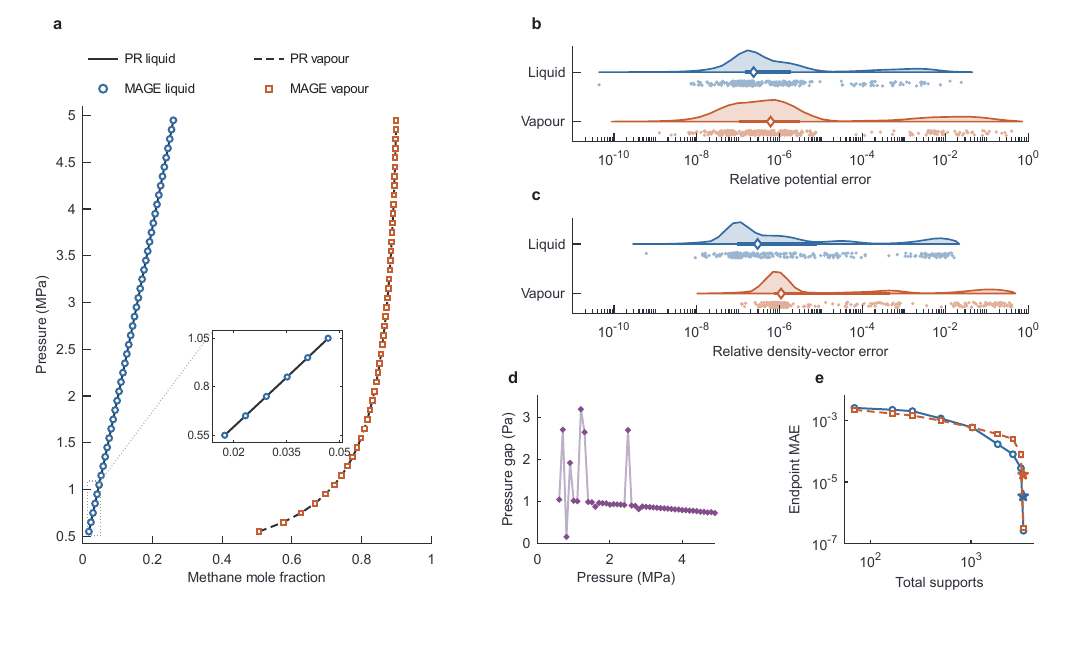}
\caption{\textbf{Binary vapour--liquid equilibrium of methane + $n$-butane at 300 K.}
\textbf{a} PR liquid and vapour coexistence curves at 45 reference pressures. Open circles
and squares mark the MAGE liquid and vapour endpoints, respectively, at each pressure. The
inset enlarges the liquid branch from 0.55 to 1.05~MPa. \textbf{b,c} Relative errors in
potential and component-density vector on 4,096 homogeneous states per branch. Half-density
profiles include all observations on a logarithmic scale. Dots show 320 states per branch,
horizontal bars span the interquartile range and open diamonds mark the medians. \textbf{d}
Absolute branch-pressure difference at the PR coexistence coordinates of the 44 midpoint
pressures. \textbf{e} Endpoint mean absolute error (MAE) over the 44 pressures. Blue circles
and orange squares give the liquid and vapour errors, respectively. Lines connect nested
support sets. Filled stars mark the selected 3,304-support model.}
\label{fig:binary}
\end{figure}

We first assess the accuracy of the generating representation for vapour--liquid equilibrium
(VLE) of methane + $n$-butane at 300~K, using the Peng--Robinson (PR) EOS\cite{peng1976}
(Methods). With 3,304 supports selected by the joint branch-gap and endpoint criteria
(Methods and Supplementary Note~3), the two branch-pressure constraints recover the reference
$P$--$x$--$y$ envelope at 45 prescribed pressures from 0.55 to 4.95~MPa, distinct from the
support pressures (Fig.~\ref{fig:binary}a). The liquid and vapour methane mole-fraction
mean absolute errors (MAEs) are $4.11\times10^{-6}$ and $8.87\times10^{-6}$, respectively.
The calculated endpoints also reproduce the PR phase labels for all 2,000 independently
sampled feeds (Supplementary Fig.~2a).

The error analysis links coexistence accuracy to the individual branch potentials and to
their difference. On 4,096 independent homogeneous states per phase, potential and
component-density errors are larger on average for vapour than for liquid
(Fig.~\ref{fig:binary}b,c). At the PR coexistence coordinates of 44 additional midpoint
pressures, however, the mean absolute difference between the MAGE branch pressures is
1.05~Pa (Fig.~\ref{fig:binary}d). The phase margin is perturbed by differences in both the
hard-envelope errors and the smoothing deficits. The latter partially cancel in the phase
margin: their mean absolute difference is below 3\% of the mean deficit of either branch.
The midpoint tests also confirm accurate endpoint recovery and prescribed-pressure closure
(Supplementary Note~3).

With coexistence accuracy established, we next assess the support count needed for a
specified endpoint accuracy. The support count determines the number of plane evaluations
in a two-branch query. Over the same 44 midpoint pressures, endpoint errors decrease along
the nested support sequence (Fig.~\ref{fig:binary}e and Supplementary Fig.~3). For an MAE
target of $10^{-3}$ in each phase, 1,029 supports suffice on this grid, requiring 69\%
fewer plane evaluations than the selected 3,304-support model, which combines branch-gap
and equilibrium criteria (Supplementary Note~3). The study thus quantifies the trade-off
between representation size and endpoint accuracy on the tested grid.

\subsection*{Nonisothermal vapour--liquid equilibrium}

Allowing the thermal coordinate $\eta_0=-1/(RT)$, which is fixed in the isothermal benchmark,
to vary extends the same generating-potential framework to coexistence and caloric properties.
For methane + $n$-butane, we combine the PR residual Helmholtz energy with an ideal-gas
contribution based on reference heat capacities and construct a single temperature-dependent
representation using the same branch-gap and endpoint criteria (Methods and Supplementary Note~3).

\begin{figure}[!htbp]
\centering
\includegraphics[width=\linewidth]{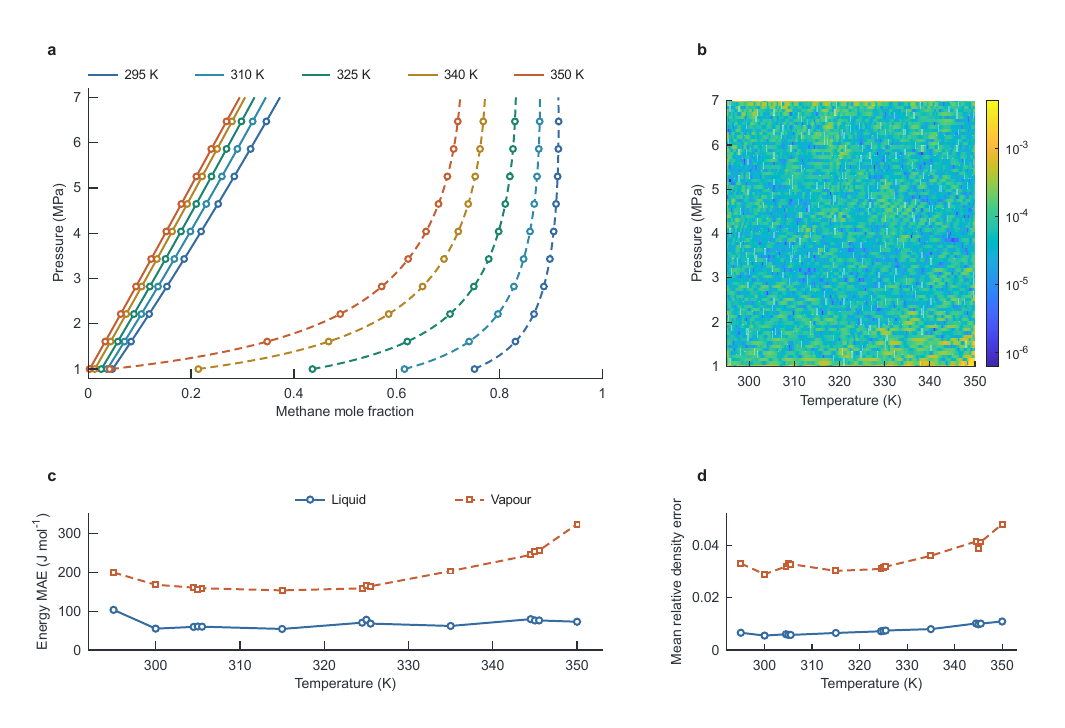}
\caption{\textbf{Nonisothermal coexistence of methane + $n$-butane.} \textbf{a} Liquid
(solid) and vapour (dashed) PR coexistence curves at five temperatures. Open circles mark
MAGE endpoints. \textbf{b} The larger of the two absolute endpoint errors in methane mole
fraction at each of the 18,204 temperature--pressure pairs. The colour scale is logarithmic.
\textbf{c,d} Mean absolute error in molar internal energy and mean relative error in the
component-density vector, grouped by test temperature for each homogeneous branch. Each
branch has 2,688 test states.}
\label{fig:noniso}
\end{figure}

This representation recovers coexistence over 18,204 temperature--pressure pairs spanning
295--350~K and 1--7~MPa. The evaluation temperatures are distinct from the support
temperatures, and the test points lie within each branch's support convex hull in the
temperature--pressure plane (Supplementary Note~3). Both liquid and vapour composition
MAEs are below $8\times10^{-5}$ (Fig.~\ref{fig:noniso}a,b), and the calculated endpoints
reproduce all phase labels in a separate sample of 2,000 feeds (Supplementary Fig.~2b).

Beyond coexistence, the derivative with respect to the thermal Massieu coordinate generates
the branch internal-energy density. On 2,688 independent homogeneous states per phase, the
liquid and vapour molar internal-energy MAEs relative to the sensible-energy reference are
69.8 and 198~J\,mol$^{-1}$, respectively (Fig.~\ref{fig:noniso}c). Across all test states
in each phase, the mean relative errors in the generated component-density vectors are
0.0076 for liquid and 0.035 for vapour (Fig.~\ref{fig:noniso}d and Supplementary Table~5).
A single temperature-dependent representation thus recovers coexistence geometry and
generates both caloric and compositional properties across the prescribed domain.

\subsection*{Reference-model compatibility and liquid--liquid coexistence}

\begin{figure}[!htbp]
\centering
\includegraphics[width=\linewidth]{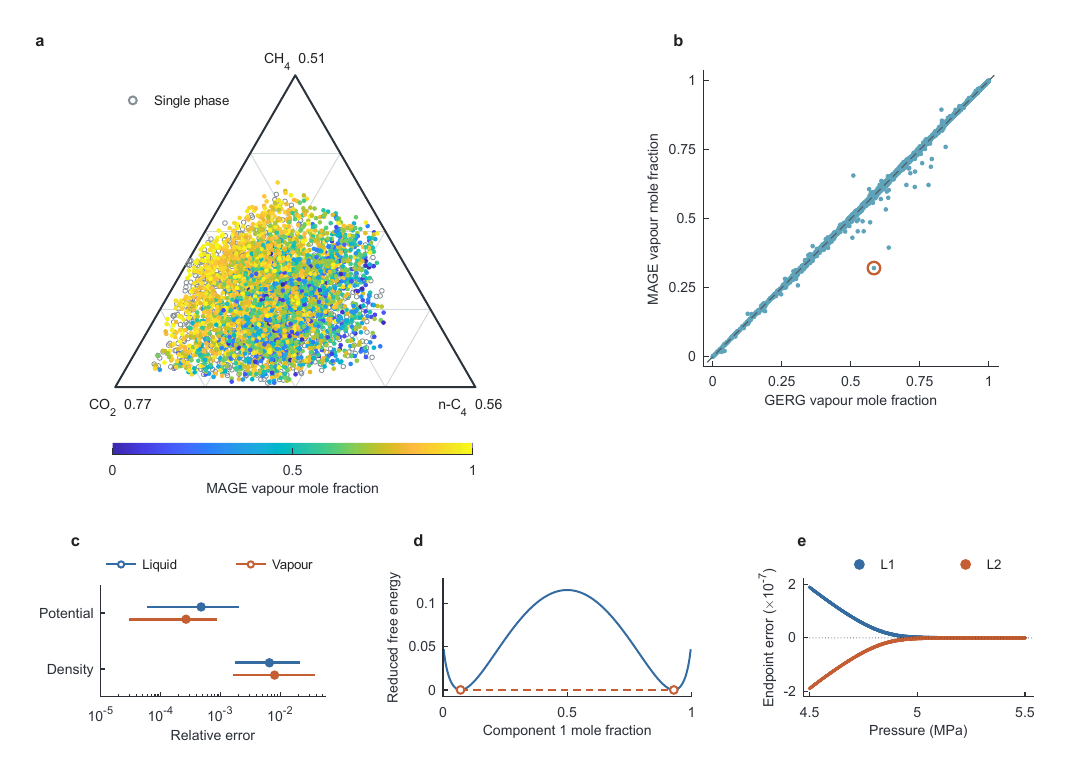}
\caption{\textbf{Phase equilibrium from multiparameter gas thermodynamics and a symmetric
liquid model.} \textbf{a} MAGE phase states on 5,000 ternary feeds at 320~K and 1.5--10~MPa,
projected onto composition. Open circles denote single-phase feeds; colour indicates the
vapour molar phase fraction of two-phase feeds. MAGE and GERG-2008 agree on the number of
equilibrium phases for every feed. The lower mole-fraction limits are 0.35 for CO$_2$,
0.09 for CH$_4$ and 0.14 for $n$-butane. Vertex labels give the upper limits. \textbf{b}
Vapour molar phase fractions from both models for the 3,963 two-phase feeds. The dashed
line indicates equality. The ring marks the largest discrepancy. \textbf{c} Relative
potential and component-density errors on 20,000 homogeneous states per branch. Dots mark
medians and bars span the 5th to 95th percentiles. \textbf{d} Symmetric liquid free energy
and common tangent. \textbf{e} Signed coexistence-composition errors (MAGE minus analytic)
for both liquids at 2,000 feed pressures.}
\label{fig:generality}
\end{figure}

Because the finite-support construction takes native thermodynamic states as inputs, it
can accommodate reference models of different functional forms; we first test this with
GERG-2008\cite{kunz2012}, a multiparameter EOS, for CO$_2$ + CH$_4$ + $n$-butane at 320~K.
The fixed 4,990-support representation agrees with GERG on the number of equilibrium phases
for all 5,000 independent confirmatory feeds at 1.5--10~MPa (Fig.~\ref{fig:generality}a).
Among the 3,963 two-phase feeds, the vapour molar phase-fraction MAE is 0.003031 and the
95th-percentile absolute error is 0.00898, with a maximum of 0.26256 (Fig.~\ref{fig:generality}b).
All feeds satisfy the pressure, composition and dual-feasibility tolerances. Independent
homogeneous tests give mean relative potential errors below $10^{-3}$ and mean relative
density-vector errors of approximately 1\% in both branches (Fig.~\ref{fig:generality}c
and Supplementary Note~3).

The largest phase-fraction discrepancy illustrates the role of coexistence sensitivity
alongside average branchwise accuracy. At 7.792~MPa, the feed $n$-butane mole fraction
lies between the phase endpoints in both models, yet the predicted and reference vapour
fractions are 0.321 and 0.584, respectively. At the GERG coexistence coordinates, the MAGE
liquid and vapour pressure errors are $-36.0$ and $-0.545$~kPa. The individual pressure
errors perturb the prescribed-pressure constraints, and their difference perturbs phase
equality. The displacement of the common coexistence coordinates contributes $-0.05267$
to the liquid $n$-butane mole-fraction error, compared with $+0.00509$ from the local
gradient error. This displacement dominates the liquid endpoint error and shortens the
predicted tie line; the short separation amplifies the phase-fraction error through the
lever rule (Supplementary Note~3). The scaled residual norm of $3.3\times10^{-16}$ at the
MAGE solution confirms numerical closure of the represented equations, while the endpoint-error
decomposition identifies how representation errors propagate into phase partitioning
(Supplementary Note~3).

Gas--brine equilibrium illustrates how the finite-support construction separates the
complexity of the reference thermodynamics from the cost of online branch evaluation.
Models of gas--brine systems, such as those used for underground hydrogen storage and
CO$_2$--brine equilibrium\cite{ahmed2024,xiong2023}, can require several coupled thermodynamic
contributions. For hydrogen--water--sodium chloride, a single Gibbs-energy function combines
Helmholtz EOSs for water and hydrogen with an aqueous activity model and hydrogen-solubility
correlations (Supplementary Note~5.1). MAGE incorporates these contributions into the
support states offline and evaluates the resulting potentials online with the same plane
sums and weighted moments as the EOS-based representations: for a branch with $M$ supports
and $d$ independent dual coordinates, evaluating the potential and gradient requires $O(Md)$
arithmetic work, and evaluating a dense Hessian requires $O(Md^2)$. At fixed $M$ and $d$,
the arithmetic cost of evaluating a branch potential and its derivatives is therefore
independent of the internal complexity of the reference model. With this common evaluation
form, MAGE reproduces all reference phase labels in an independent confirmation sample of
3,584 feeds, with a vapour-fraction MAE of $1.00\times10^{-4}$ (Supplementary Table~11).

The same phase-potential construction also describes competition between distinct liquid
phases. In a symmetric binary Helmholtz model at 300~K, the two branches recover the
analytic common-tangent compositions with a composition MAE of $3.29\times10^{-8}$ for
each liquid and reproduce all phase labels for 2,000 random feeds (Methods;
Fig.~\ref{fig:generality}d,e). Along the coexisting chemical-potential line, the analytic
curvature gives a potential-error bound of 0.0675~mol\,m$^{-3}$, which exceeds the measured
maximum error of 0.0364~mol\,m$^{-3}$ (Supplementary Note~2). The liquid--liquid benchmark
thus connects the dual representation to classical common-tangent geometry and tests the
potential-error bound.

\subsection*{Phase appearance and disappearance in pressure-coupled ternary flow}

\begin{figure}[!htbp]
\centering
\includegraphics[width=\linewidth]{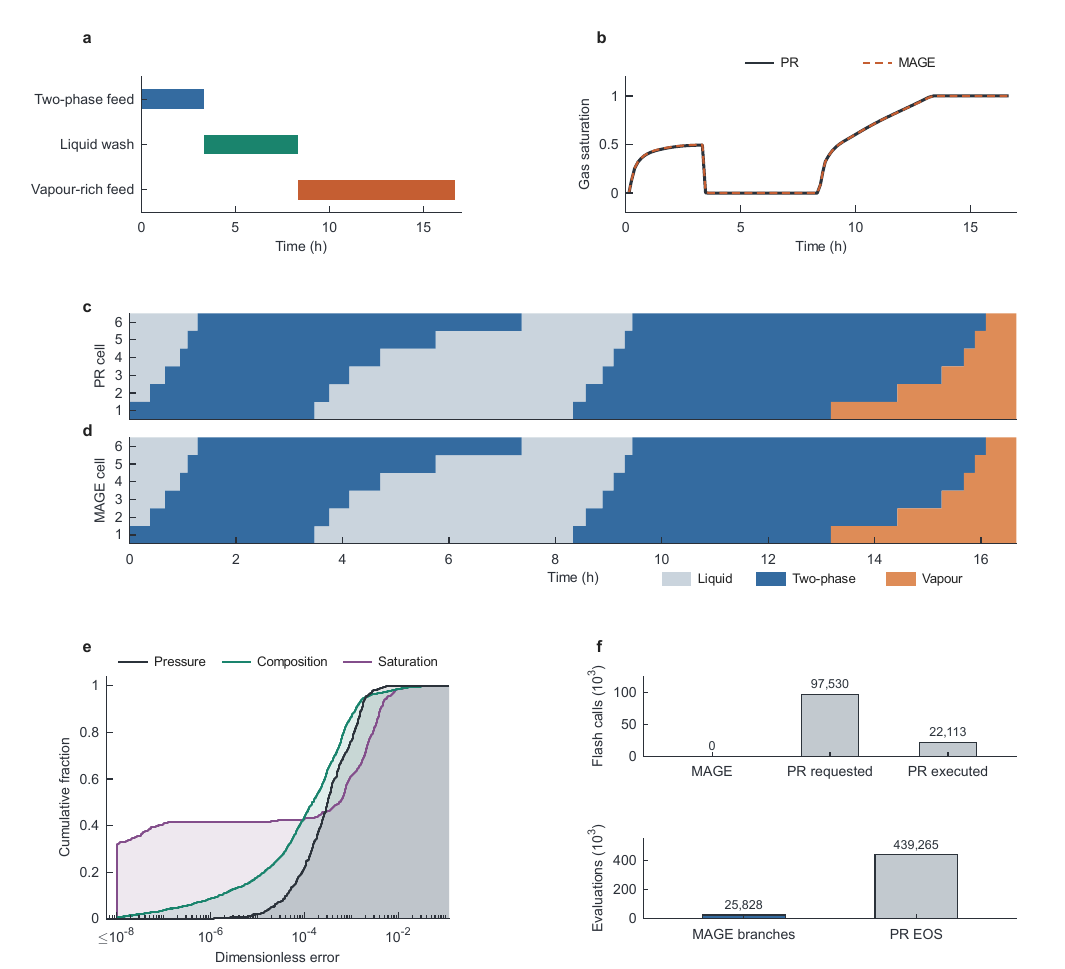}
\caption{\textbf{Pressure-coupled ternary flow with repeated phase appearance and
disappearance.} \textbf{a} Injection sequence at 320~K. \textbf{b} Gas saturation in cell
1. \textbf{c,d} Phase state by cell and time for PR and MAGE on the same 92 accepted steps.
Each interval shows the phase label at the end of the step. \textbf{e} Cumulative distributions
of relative pressure error, absolute overall mole-fraction error and absolute saturation
error. Pressure and saturation errors each comprise 552 cell--time states; composition
errors are pooled across all three components. The leftmost tick includes all errors of
$10^{-8}$ or less. \textbf{f} Operation counts over the 92 time steps. The upper bars show
MAGE nested-flash calls, PR flash requests during global residual evaluation and PR flashes
executed during integration. The lower bars count MAGE branch evaluations and PR
equation-of-state evaluations. Numbers above the bars give the totals.}
\label{fig:coupled}
\end{figure}

Having established stand-alone equilibrium accuracy, we test the generating potentials as
a direct thermodynamic closure within the global conservation problem. We simulate the
CH$_4$ + CO$_2$ + $n$-butane mixture at 320~K in a one-dimensional six-cell finite-volume
domain, with inlet and outlet pressures of 3.2 and 2.0~MPa. A three-stage injection sequence
drives the mixture repeatedly across phase boundaries: 12,000~s of two-phase-feed injection,
18,000~s of liquid wash and 30,000~s of vapour-rich injection (Fig.~\ref{fig:coupled}a).
Both solvers use the same fixed sequence of 92 nonuniform, fully implicit time steps.

With fixed support sets, MAGE determines five global unknowns per cell,
$(\eta_{\mathrm{CH_4}},\allowbreak\eta_{\mathrm{CO_2}},\allowbreak\eta_{\mathrm{nC_4}},\allowbreak
S_g,\allowbreak P)$, from three component balances, pressure closure and bounded phase
complementarity. The independent PR reference instead uses pressure and two overall-composition
coordinates, with local flash calculations supplying phase states. Nine PR flashes initialize
the MAGE inverse solves before time integration. Supplementary Notes~3 and 4 specify the
physical model, discrete equations, support construction and reference-equilibrium checks.

The injection sequence produces four phase transitions in every cell. In cell 1, the first
injection produces a two-phase state, the wash restores a liquid state, and the final
injection produces a second two-phase state followed by a vapour state (Fig.~\ref{fig:coupled}b).
The corresponding fronts propagate from the inlet to the outlet in both solutions
(Fig.~\ref{fig:coupled}c,d). The two formulations agree on phase state at all 552 paired
cell--time states and on all 24 independently identified events in time step, cell and
transition type. Continuous saturation and branch-pressure complementarity therefore recover
the observed sequence of phase appearance and disappearance.

The agreement in phase evolution extends to the pressure, composition and saturation fields.
The mean relative pressure error is below $10^{-3}$, and the MAEs in overall mole fraction
and gas saturation are $7.70\times10^{-4}$ and $1.52\times10^{-3}$, respectively
(Fig.~\ref{fig:coupled}e). The largest mole-fraction error, 0.0777, occurs for $n$-butane
in the outlet cell during passage of the final front; both solutions transition from two
phases to vapour at the next step. MAGE maintains component conservation, with a maximum
relative component mass imbalance of $4.90\times10^{-13}$. The complete error statistics,
solver residuals and PR fugacity, material-balance and multistart stability checks are
given in Supplementary Note~4.

Operation counts during integration show how this closure reorganizes the thermodynamic
computation. MAGE evaluates the fixed branch potentials and their derivatives directly,
with zero nested local flashes across the 92 steps; the PR reference executes 22,113
flashes, including accepted-state evaluations (Fig.~\ref{fig:coupled}f). Supplementary
Note~4 reports the complete operation counts and elapsed times of the two implementations.

Shared chemical Massieu coordinates, branch-pressure complementarity and material balance
resolve phase transitions within the global conservation problem, with branch gradients
supplying densities for accumulation and fluxes. The coupled-flow results establish that,
within the prescribed state domain, thermodynamic equilibrium can be organized as a reusable
generating-potential representation that directly closes conservation equations.

\section*{Discussion}

The reusable generating-potential representation links the convex geometry of equilibrium
to the global nonlinear problem through state generation and phase competition in shared
intensive coordinates. The subdifferential of the equilibrium potential describes mixture
states, and the prescribed-pressure dual optimality conditions give material balance and
phase complementarity. These relations make constitutive evaluation and phase transitions
part of the conservation problem itself. More generally, directly representing a common
mathematical structure can change how the local computations that recover it are organized
and coupled.

Equilibrium accuracy depends jointly on the branch potentials, which enter the pressure
constraints, on their differences, which govern phase competition, and on their gradients,
which supply the phase densities. Coverage and smoothing control all three quantities
(Supplementary Note~2), whereas the sensitivity of the equilibrium state determines how
their errors propagate, as the GERG diagnostic illustrates through coordinate displacement
and tie-line shortening. Closure residuals measure how accurately the represented equilibrium
is solved, whereas comparison with the reference thermodynamics measures the fidelity of
that representation. For phase-equilibrium surrogates and other representations built from
competing generating functions, this suggests a design principle: function values, gradients
and competition margins should be assessed and controlled jointly, taking account of the
sensitivity of the resulting physical state, especially near degeneracy.

Native thermodynamic states provide a common interface between reference models and online
evaluation. Cubic, multiparameter Helmholtz and composite Gibbs-energy models all supply
inputs to the same finite-support construction, and the resulting potentials share an
evaluation form whose arithmetic cost per branch evaluation is set by the support count
and coordinate dimension. More elaborate reference thermodynamics can thus be incorporated
offline at a given representation size and coordinate dimension, and the nonisothermal
formulation extends the interface to energy as well as component densities. Reuse across
injection histories within the prescribed domain (Supplementary Note~5.2) and evaluation
timings on a fixed two-parameter, six-component state family (Supplementary Note~6) further
illustrate this separation of reference-model evaluation from downstream computation. Such
structure-preserving interfaces offer a route to organizing reference modelling, thermodynamic
representation and coupled computation as connected computational layers.

Support coverage, curvature and smoothing determine approximation accuracy. Under the
stated regularity conditions, uncalibrated potentials converge uniformly as fill distance
and smoothing bound vanish; generated states converge uniformly under the gradient-bound
assumptions. Convergence extends to regular coexistence and nondegenerate partitioning
under the corresponding conditions, including vanishing calibration for prescribed pressure
(Supplementary Note~2). The isometric embedding control illustrates the role of intrinsic
domain geometry in coverage (Supplementary Note~7). Two connected directions follow. Support
placement, selected here by a greedy rule that combines coverage, phase-margin drift and
calibration, could be optimized against domain geometry, curvature and equilibrium sensitivity
to balance representation size, accuracy and evaluation cost. Smoothing also shapes the
branch Hessian and influences nonlinear iteration counts (Supplementary Note~4). Joint
adaptation of support placement and smoothing to equilibrium sensitivity and nonlinear
solver behaviour could yield more compact, accurate and efficient thermodynamic representations.

\section*{Methods}

\subsection*{Thermodynamic formulation}

For a nonreactive multicomponent system, the fundamental thermodynamic relation reads
\begin{equation}
\mathrm{d} U=T\,\mathrm{d} S-p\,\mathrm{d} V+\sum_i\mu_i\,\mathrm{d} N_i.
\label{eq:fundamental}
\end{equation}
With $u=U/V$, $s=S/V$ and $c_i=N_i/V$, extensivity gives
\begin{equation}
\mathrm{d} u=T\,\mathrm{d} s+\sum_i\mu_i\,\mathrm{d} c_i.
\end{equation}
We define the convex reduced negative entropy density
\begin{equation}
h(\boldsymbol q)=-\frac{s(\boldsymbol q)}{R},
\end{equation}
so that
\begin{equation}
\mathrm{d} h=-\frac{1}{RT}\,\mathrm{d} u+\sum_i\frac{\mu_i}{RT}\,\mathrm{d} c_i
=\boldsymbol\eta^{\mathsf T}\mathrm{d}\boldsymbol q,
\end{equation}
which identifies the Massieu coordinates of Eq.~\eqref{eq:eta}. The Legendre--Fenchel conjugate of
$h$ is
\begin{equation}
\Phi(\boldsymbol\eta)=h^*(\boldsymbol\eta)=\sup_{\boldsymbol q}
\left[\boldsymbol\eta^{\mathsf T}\boldsymbol q-h(\boldsymbol q)\right].
\label{eq:fenchel}
\end{equation}
Combining Eq.~\eqref{eq:fenchel} with the Euler relation $u=Ts-p+\sum_i\mu_i c_i$ gives
\begin{equation}
\Phi=\frac{p}{RT},
\label{eq:phi_pressure}
\end{equation}
and, wherever $\Phi$ is differentiable, $\nabla_{\boldsymbol\eta}\Phi=\boldsymbol q$. Temperature,
chemical potentials, pressure and entropy are recovered as
\begin{equation}
T=-\frac{1}{R\eta_0},\qquad
\mu_i=-\frac{\eta_i}{\eta_0},\qquad
p=RT\Phi,\qquad
s=R(\Phi-\boldsymbol\eta^{\mathsf T}\boldsymbol q).
\end{equation}
At fixed $\boldsymbol\eta$, equilibrium among the represented homogeneous phases $\alpha$
corresponds to the maximum of $\Phi_\alpha$. At coexistence, the active branches have equal
$\Phi$ and therefore equal pressure at
the common $T$ and $\mu_i$, and the subdifferential of the equilibrium potential is the convex hull
of the coexisting density states. The full derivation and units are given in Supplementary Note~1.

\subsection*{Finite-support Massieu approximation}

For phase $\alpha$, the Massieu potential is
\begin{equation}
\Phi_\alpha(\boldsymbol\eta)=\sup_{\boldsymbol q\in\mathcal D_\alpha}
\left[\boldsymbol\eta^{\mathsf T}\boldsymbol q-h_\alpha(\boldsymbol q)\right],
\end{equation}
where $\mathcal D_\alpha$ denotes the represented homogeneous phase domain. The restricted
conjugate is convex as a supremum of affine functions of $\boldsymbol\eta$. For any native
state in $\mathcal D_\alpha$, the affine function $\boldsymbol\eta^{\mathsf T}\boldsymbol
q_m-h_\alpha(\boldsymbol q_m)$ is a lower bound on $\Phi_\alpha$. Attainment of the supremum
at its native coordinate gives a contact tangent; at a differentiable point its density
equals the branch gradient. A stationary interior homogeneous state is a contact state
when the reduced negative entropy is differentiable and convex on a convex branch domain,
including for metastable states on a restricted branch. The Bregman and curvature bounds
use contact tangents with the regularity specified in Supplementary Notes~1 and 2. Native
homogeneous branch states and coexistence data guide the construction of the material-specific
tangent library. The maximum of its planes defines the hard envelope. For a common additive
offset $\Delta\Phi$, the normalized log-sum-exp satisfies

\begin{equation}
\max_m L_{\alpha m}-\tau\log M
\le \widehat\Phi_\alpha-\Delta\Phi\le \max_m L_{\alpha m}.
\end{equation}
More precisely, the deficit of the smoothed branch relative to the hard envelope depends
on the local arrangement of the supports,
\begin{equation}
\widehat\Phi_\alpha-\Delta\Phi=\max_m L_{\alpha m}-\tau\log\frac{M}{N_{\mathrm{eff},\alpha}},\qquad
N_{\mathrm{eff},\alpha}=\sum_{m=1}^{M}\exp\left(\frac{L_{\alpha m}-\max_j L_{\alpha
j}}{\tau}\right),
\label{eq:deficit}
\end{equation}
where $1\le N_{\mathrm{eff},\alpha}\le M$ is an effective count of the supports whose
planes lie within approximately $\tau$ of the dominant plane. Equations~\eqref{eq:gradmage}
and \eqref{eq:hessmage}
follow by differentiation; the Hessian is positive definite on the independent coordinates
when the projected support slopes affinely span that space and otherwise has null directions.
Because an exact tangent support differs from the exact branch by its
Bregman divergence, bounds on the curvature and on the support fill distance provide an a priori
bound on the hard-envelope error, and Eq.~\eqref{eq:deficit} provides the smoothing contribution.
Supplementary Note~2 develops these bounds and their propagation to state gradients, phase margins,
coexistence coordinates and phase compositions. A shared additive calibration shifts
pressure by $RT\Delta\Phi$ while preserving branch differences and derivatives at fixed
dual coordinates. For a prescribed physical pressure, the calibration enters the coexistence
equations and therefore affects the recovered coordinates and phase compositions. In the
coupled-flow representation, $\Delta\Phi=3.80561905$~mol\,m$^{-3}$ is the mean potential
deficit of the offline candidate pool used to construct the input library. It enters both
branch pressures and their saturation-weighted pressure closure, giving $RT\Delta\Phi=10.13$~kPa
at 320~K. Supplementary Note~3 gives the offsets for each material.

For numerical construction, coordinates and potentials are scaled using a reference Massieu
coordinate $\boldsymbol\eta_{\rm ref}$, a reference potential $\Phi_{\rm ref}$, a diagonal
coordinate-scale matrix $D_\eta$ and a positive potential scale $\Phi_s$:
\begin{equation}
\boldsymbol X=D_\eta^{-1}(\boldsymbol\eta-\boldsymbol\eta_{\rm ref}),\qquad
\varphi=\frac{\Phi-\Phi_{\rm ref}}{\Phi_s},
\end{equation}
so that the support slopes become $\boldsymbol g_m=D_\eta^{\mathsf T}\boldsymbol q_m/\Phi_s$ and the
physical gradients are recovered as $\boldsymbol q=\Phi_sD_\eta^{-\mathsf
T}\nabla_X\widehat\varphi$. The potentials and pressures of competing branches are evaluated
in the same physical coordinates. Branchwise standard deviations set the distance scale
for selecting candidates.

For each material, offline construction uses one library of native tangents per branch,
together with calibration states and separate tuning states. At a common probe
$\boldsymbol\eta_j$, let $\mathcal L$ denote the pair of complete libraries and $\mathcal
S$ the retained subsets. The change in the potential difference is
\begin{equation}
 d_j=(\widehat\Phi_{1,\mathcal S}-\widehat\Phi_{2,\mathcal S})(\boldsymbol\eta_j)
       -(\widehat\Phi_{1,\mathcal L}-\widehat\Phi_{2,\mathcal L})(\boldsymbol\eta_j).
 \label{eq:selection_gap}
\end{equation}
The drift $d_j$ isolates the effect of support selection at fixed scales. The total deviation
of the selected phase margin from the reference is the sum of this drift and the
complete-library approximation error (Supplementary Note~2). Candidates are stratified by
proximity to calibration states in scaled dual coordinates. Farthest-point sampling within
the strata gives initial coverage of the two branches. The selector ranks unused planes
by the marginal reduction in $|d_j|$ at probes where the current drift is largest. The
subsets must also meet a thermodynamic calibration criterion appropriate to the state
specification, such as the displacement of the coexistence endpoints or the phase stability
and feasibility of a pressure--composition solution.

The selector then tests deletions of planes with small maximum normalized weight on the
probes. A deletion is accepted when both the potential-gap and physical criteria are
satisfied. Supplementary Table~3 lists the material-specific gap tolerances. The 99th
percentile of $|d_j|$ must lie within the prescribed tolerance and its maximum within six
times that value. The equilibrium checks compare the subset with the complete library on
calibration feeds or coexistence probes. Separate tuning states determine a common multiplier
for the two branch smoothing scales. The selected planes, scales and shared offset are
frozen for online evaluation. Supplementary Note~3 specifies the partition, tolerances and
acceptance rules.

\subsection*{Reference thermodynamics and equilibrium evaluation}

In the binary and ternary PR benchmarks, all reference properties are derived consistently
from the same equation of state. Both benchmarks use classical one-fluid mixing rules with
$a_{ij}=(a_i a_j)^{1/2}(1-k_{ij})$, $a_{\rm mix}=\sum_i\sum_j x_i x_j a_{ij}$, $b_{\rm
mix}=\sum_i x_i b_i$, and $k_{ij}=0$ for all component pairs. Liquid and vapour supports
are evaluated on their respective homogeneous
phase branches, and the pressure, chemical potentials, internal-energy density and component molar
densities of each support are obtained from a single, internally consistent reference state. The
support planes of Eq.~\eqref{eq:support} are affine lower bounds on the restricted conjugates;
the contact and curvature conditions are described in Supplementary Notes~1--3. For the
nonisothermal methane--$n$-butane calculation, a
concentration-based ideal-gas Helmholtz contribution is added to the PR residual Helmholtz energy:
\begin{equation}
\begin{aligned}
 f^{\mathrm{id}}(T,\boldsymbol c)&=RT\sum_i
c_i\left[\ln\!\left(\frac{c_iRT}{p^\circ}\right)-1\right]\\
 &\quad+\sum_i c_i g_i^\circ(T),\\
 g_i^\circ(T)&=h_i^\circ(T)-Ts_i^\circ(T).
\end{aligned}
 \label{eq:caloric_ideal}
\end{equation}
The ideal-gas heat capacities are represented by Planck--Einstein correlations derived from the
methane\cite{setzmann1991} and $n$-butane\cite{buecker2006} reference equations of state.

The thermodynamic reference state is $T^\circ=298.15$~K and $p^\circ=10^5$~Pa, with
$h_i^\circ(T^\circ)=0$ and $s_i^\circ(T^\circ,p^\circ)=0$ for each species. These values define
sensible ideal-gas origins for enthalpy and entropy. The resulting internal-energy reference
preserves PR pressures and vapour--liquid equilibrium conditions. Pressure, chemical
potentials and internal energy are obtained from derivatives of the same total Helmholtz
energy. The correlations,
coefficients and analytic caloric integrals are specified in Supplementary Note~3.

For binary stand-alone VLE at specified $T$ and $p$, $\eta_0$ is fixed and the chemical
Massieu coordinates
are obtained by solving
\begin{equation}
\widehat\Phi_l(\boldsymbol\eta)=\widehat\Phi_g(\boldsymbol\eta)=\frac{p}{RT}.
\end{equation}
The coexisting phase compositions then follow from the branch gradients (Eq.~\eqref{eq:gradmage}).
The binary support-count studies use identical candidate, validation and test pools for every $M$.

The GERG benchmark uses GERG-2008\cite{kunz2012} as the fundamental Helmholtz-energy reference.
Properties of the CH$_4$ + CO$_2$ + $n$-butane mixture are evaluated with the Python interface of
Thermopack version 2.2.3 (SINTEF Energy Research and NTNU)\cite{wilhelmsen2017}, using the
software's default reference state. For each $(T,p,\boldsymbol z)$, the two MAGE branches
supply pressures $p_\alpha=RT\widehat\Phi_\alpha$ and concentrations $\boldsymbol
c^\alpha=\nabla_{\eta_c}\widehat\Phi_\alpha$. Nonnegative phase volumes satisfy composition
balance and complementarity with $p-p_\alpha$. Each occupied phase reaches the prescribed
pressure. The vapour molar phase fraction follows from the phase volumes and molar densities.
We compare each resulting state with a separate TP flash and recompute its residuals
directly from the stored planes (Supplementary Note~3).

The liquid--liquid benchmark uses the Helmholtz-energy density
\begin{equation}
f(T,c_1,c_2)=\frac{K}{2}\left(\frac{c-c_0}{c_0}\right)^2-p_0
+RTc\left[x\ln x+(1-x)\ln(1-x)+\chi x(1-x)\right],
\end{equation}
where $c=c_1+c_2$, $x=c_1/c$, $c_0=10^4$ mol\,m$^{-3}$, $K=10^9$~Pa, $p_0=5$~MPa and $\chi=3$. The
analytic coexistence compositions satisfy $\ln[x/(1-x)]+\chi(1-2x)=0$. The pressure obtained
from $p=\sum_i c_i\mu_i-f$ is $p_0+(K/2)[(c/c_0)^2-1]$.

\subsection*{Coupled multiphase formulation and numerical procedures}

The coupled benchmark is isothermal, so $\eta_0$ is fixed and the chemical coordinate vector is
$\boldsymbol\eta_c=(\eta_1,\ldots,\eta_{N_{\mathrm c}})^{\mathsf T}$. For phase $\alpha\in\{l,g\}$,
\begin{equation}
p_\alpha(\boldsymbol\eta_c)=RT\widehat\Phi_\alpha,
\qquad
c_i^\alpha(\boldsymbol\eta_c)=\frac{\partial\widehat\Phi_\alpha}{\partial\eta_i}.
\end{equation}
With porosity $\phi$ and gas saturation $S_g$, the component inventory per unit bulk volume is
\begin{equation}
C_i=\phi\left[(1-S_g)c_i^l+S_g c_i^g\right],
\label{eq:inventory}
\end{equation}
and, neglecting diffusion, gravity and capillary pressure, component conservation reads
\begin{equation}
\frac{\partial C_i}{\partial t}
+\nabla\cdot\left(c_i^l\boldsymbol u_l+c_i^g\boldsymbol u_g\right)=q_i,
\end{equation}
with the Darcy velocities
\begin{equation}
\boldsymbol u_\alpha=-\boldsymbol K\frac{k_{r\alpha}}{\zeta_\alpha}\nabla P.
\end{equation}
Here $q_i$ is the component source term, $\boldsymbol K$ is the absolute permeability, $k_{r\alpha}$
is the relative permeability and $\zeta_\alpha$ is the phase viscosity. Pressure is a global unknown
and is linked to the phase branches through
\begin{equation}
R_P=P-\left[(1-S_g)p_l+S_gp_g\right]=0.
\label{eq:pclosure}
\end{equation}
Phase existence is expressed through the complementarity condition
\begin{equation}
R_{\rm ph}=S_g-
\Pi_{[0,1]}\left[S_g+\frac{p_g-p_l}{P_{\rm scale}}\right]=0,
\label{eq:phasecomp}
\end{equation}
with $P_{\rm scale}=1$~MPa. The exact projection gives $S_g=0$ when the liquid branch has
the larger pressure, $0<S_g<1$ only when $p_l=p_g$, and $S_g=1$ when the vapour branch has
the larger pressure. The global solver uses the piecewise
generalized derivative of the projection in the analytic Jacobian. A single phase residual,
depending on saturation and branch pressures, applies in all phase regimes throughout the time step.

We discretize the conservation equations with cell-centred finite volumes and fully implicit time
integration. The MAGE unknown vector in each cell is
$(\eta_{\mathrm{CH_4}},\eta_{\mathrm{CO_2}},\eta_{\mathrm{nC_4}},S_g,P)$, and component fluxes are
computed with upwind phase properties and Darcy mobilities. The fully implicit PR reference uses
pressure and two unconstrained overall-composition coordinates per cell, mapped onto the composition
simplex, and performs a local PR flash inside every global residual evaluation. This flash combines
analytic cubic roots, $K$-value initialization from the Wilson correlation or the previous time
step, the Rachford--Rice equation and successive substitution on the fugacity equations, and assigns
the phase state by comparing the Gibbs energy of the two-phase candidate with that of the
homogeneous root of lowest Gibbs energy (Supplementary Note~4). Exact-state caching reuses
numerically identical flash results during finite-difference Jacobian evaluations without
altering the mathematical formulation. Both solvers use bound-constrained trust-region
least-squares iterations with identical physical parameters, step, cost and gradient tolerances of
$10^{-11}$, a common residual acceptance threshold of $10^{-6}$ and a common fixed time grid (92
steps for the 12,000~s first injection). MAGE supplies analytic phase-branch and generalized
complementarity
derivatives to its global Jacobian, whereas the PR reference uses finite-difference global
Jacobians. The complete discrete residuals, bounds, convergence criteria, phase-label convention and
time grid are given in Supplementary Note~4.

The relative pressure error, absolute overall mole-fraction error and absolute gas-saturation error
are evaluated over all 552 common cell--time states in the schedule with a 12,000~s first
injection. Phase labels are
assigned only during post-processing, using $S_g<10^{-4}$ for liquid, $S_g>1-10^{-4}$ for vapour and
two-phase otherwise. The online workload counts refer to time integration. The paired
trajectory timings include boundary and initial-state mapping; the first-use timings
additionally include offline support construction (Supplementary Note~4).

\section*{Data availability}

The numerical inputs, selected support states, independent reference
states, equilibrium solutions, flow histories and figure source
tables are publicly available on Zenodo at
\url{https://doi.org/10.5281/zenodo.24095790}.
The archive includes the numerical data underlying the main figures
and supplementary tables in their native numerical formats.

\section*{Code availability}

The source code, frozen finite-support Massieu models, reference
thermodynamic implementations, equilibrium solvers and figure-generation routines are publicly available on GitHub at
\url{https://github.com/PiyangLiu/MAGE_PROJECT}.
The version used in this study is archived on Zenodo at
\url{https://doi.org/10.5281/zenodo.24097456}.

\section*{Acknowledgements}

\begingroup\raggedright
This work was supported by the National Natural Science Foundation of China (Grant Nos. 52374036,
52674045, 52441411) and the Shandong Provincial Natural Science Foundation, China (Grant No.
ZR2023ME207).\par\endgroup

\section*{Author contributions}

P.L.: Conceptualization, Methodology, Software, Investigation, Formal analysis, Visualization,
Writing -- original draft. K.Z.: Methodology, Supervision, Project administration. S.S.: Software,
Validation, Data curation, Investigation, Writing -- review and editing. X.Y.: Methodology, Formal
analysis, Validation. J.Y.: Conceptualization, Supervision, Resources.

\section*{Competing interests}

The authors declare no competing interests.

\end{document}


\maketitle

\section*{Supplementary Note 1: Massieu--Fenchel thermodynamic derivation}

The dual coordinates and the generating potential follow from the fundamental thermodynamic
relation for an extensive, nonreactive multicomponent system:
\begin{equation}
\mathrm{d} U=T\,\mathrm{d} S-p\,\mathrm{d} V+\sum_i\mu_i\,\mathrm{d} N_i.
\end{equation}
The corresponding densities are defined as
\begin{equation}
u=U/V,\qquad s=S/V,\qquad c_i=N_i/V,
\end{equation}
with units of J\,m$^{-3}$, J\,m$^{-3}$\,K$^{-1}$ and mol\,m$^{-3}$, respectively. Euler homogeneity
gives
\begin{equation}
u=Ts-p+\sum_i\mu_i c_i,
\label{eq:SIeuler}
\end{equation}
and the differential density relation
\begin{equation}
\mathrm{d} u=T\,\mathrm{d} s+\sum_i\mu_i\,\mathrm{d} c_i.
\end{equation}
We define
\begin{equation}
\boldsymbol q=(u,c_1,\ldots,c_{N_{\mathrm c}})^{\mathsf T},\qquad
h(\boldsymbol q)=-\frac{s(\boldsymbol q)}{R}.
\end{equation}
The reduced negative entropy density $h$ has units of mol\,m$^{-3}$ and is convex on a stable
homogeneous thermodynamic branch. Its differential is
\begin{equation}
\mathrm{d} h=-\frac{1}{RT}\,\mathrm{d} u+\sum_i\frac{\mu_i}{RT}\,\mathrm{d} c_i
=\boldsymbol\eta^{\mathsf T}\mathrm{d}\boldsymbol q,
\end{equation}
where
\begin{equation}
\eta_0=-\frac{1}{RT},\qquad
\eta_i=\frac{\mu_i}{RT}.
\end{equation}
Thus, $\eta_0$ has units of mol\,J$^{-1}$, whereas each chemical coordinate $\eta_i$ is
dimensionless.

The Legendre--Fenchel conjugate of $h$ is\cite{rockafellar1970}
\begin{equation}
\Phi(\boldsymbol\eta)=h^*(\boldsymbol\eta)=
\sup_{\boldsymbol q}\left[\boldsymbol\eta^{\mathsf T}\boldsymbol q-h(\boldsymbol q)\right].
\label{eq:SIconjugate}
\end{equation}
At conjugate states, the Fenchel--Young equality holds:
\begin{equation}
h(\boldsymbol q)+\Phi(\boldsymbol\eta)=\boldsymbol\eta^{\mathsf T}\boldsymbol q.
\end{equation}
Substituting the definitions of $h$ and $\boldsymbol\eta$ and using Eq.~\eqref{eq:SIeuler} yields
\begin{align}
\Phi
&=\frac{s}{R}-\frac{u}{RT}+\sum_i\frac{\mu_i c_i}{RT}\\
&=\frac{p}{RT}.
\end{align}
Hence, $\Phi$ has units of mol\,m$^{-3}$. Where the conjugate is differentiable,
\begin{equation}
\frac{\partial\Phi}{\partial\eta_0}=u,\qquad
\frac{\partial\Phi}{\partial\eta_i}=c_i,
\end{equation}
that is, $\nabla_{\boldsymbol\eta}\Phi=\boldsymbol q$. The intensive variables and entropy
density can then be recovered as
\begin{equation}
T=-\frac{1}{R\eta_0},\qquad
\mu_i=-\frac{\eta_i}{\eta_0},\qquad
p=RT\Phi=-\frac{\Phi}{\eta_0},
\end{equation}
\begin{equation}
s=R(\Phi-\boldsymbol\eta^{\mathsf T}\boldsymbol q).
\end{equation}
The total molar density is $c=\sum_i c_i$, the mole fraction is $x_i=c_i/c$ and the mass density is
$\rho=\sum_i M_i c_i$, where $M_i$ is the molar mass of component $i$.

For phase-specific homogeneous domains $\mathcal D_\alpha$, define the extended function
$h_\alpha$ by the reduced negative entropy on $\mathcal D_\alpha$ and $+\infty$ elsewhere.
Its restricted conjugate $\Phi_\alpha=h_\alpha^*$ is convex because it is the supremum of
affine functions. For every admissible homogeneous state $\boldsymbol q_m\in\mathcal
D_\alpha$, the native affine function obeys
\begin{equation}
L_m(\boldsymbol\eta)=\boldsymbol\eta^{\mathsf T}\boldsymbol q_m-h_\alpha(\boldsymbol q_m)
\leq\Phi_\alpha(\boldsymbol\eta).
\label{eq:SIlowerSupport}
\end{equation}
At its native coordinate $\boldsymbol\eta_m=\nabla h_\alpha(\boldsymbol q_m)$ the Euler
relation gives $L_m(\boldsymbol\eta_m)=p_m/(RT_m)$. The Fenchel--Young contact condition is
\begin{equation}
\Phi_\alpha(\boldsymbol\eta_m)+h_\alpha(\boldsymbol q_m)
=\boldsymbol\eta_m^{\mathsf T}\boldsymbol q_m.
\label{eq:SIcontact}
\end{equation}
It is equivalent to attainment of the restricted supremum at $\boldsymbol q_m$, or to the
supporting inequality $h_\alpha(\boldsymbol q)\geq h_\alpha(\boldsymbol
q_m)+\boldsymbol\eta_m^{\mathsf T}(\boldsymbol q-\boldsymbol q_m)$ for every $\boldsymbol
q\in\mathcal D_\alpha$. A stationary interior state is sufficient when $h_\alpha$ is
differentiable and convex on a convex branch domain; strict convexity gives uniqueness.
At contact, $\boldsymbol q_m\in\partial\Phi_\alpha(\boldsymbol\eta_m)$, and differentiability
gives $\nabla\Phi_\alpha(\boldsymbol\eta_m)=\boldsymbol q_m$.

At fixed temperature, eliminating the energy density by the thermal Legendre transform
gives the equivalent maximization of $[\boldsymbol\mu^{\mathsf T}\boldsymbol
c-f_\alpha(T,\boldsymbol c)]/(RT)$, where $f_\alpha$ is the Helmholtz-energy density on
the restricted branch. Positive semidefiniteness of $\nabla^2_{\boldsymbol c}f_\alpha$ is
a necessary second-order condition for local stability with respect to concentration
variations; positive definiteness ensures a strict local maximum of the intensive-variable
objective at a stationary state. Convexity on a convex restricted concentration domain
yields the supporting inequality throughout that domain. Locally stable metastable states
can satisfy this condition within their own branch domain. Thus, membership in the branch
domain ensures the affine lower bound, contact makes this bound a tangent, and the
differentiability and curvature assumptions in Supplementary Note~2 yield the Bregman error
estimates.

The equilibrium potential of the represented phase family at fixed $\boldsymbol\eta$ is
the maximum of the Massieu potentials,
\begin{equation}
\Phi_{\rm eq}(\boldsymbol\eta)=\max_\alpha\Phi_\alpha(\boldsymbol\eta).
\end{equation}
Two active phases coexist when $\Phi_\alpha=\Phi_\beta=\Phi_{\rm eq}$. Because the same
$\boldsymbol\eta$ fixes
$T$ and all $\mu_i$, equality of the active Massieu potentials is equivalent to equality of
pressure. At coexistence, the equilibrium potential is generally nondifferentiable, and
\begin{equation}
\partial\Phi_{\rm eq}
=\operatorname{conv}\{\nabla\Phi_\alpha:\alpha\ \text{active}\}
=\operatorname{conv}\{\boldsymbol q^\alpha\}.
\end{equation}
A bulk mixture state can therefore be written as $\boldsymbol q_{\rm mix}=\sum_\alpha
S_\alpha\boldsymbol q^\alpha$, where $S_\alpha\ge0$ is the volume fraction of phase $\alpha$ and
$\sum_\alpha S_\alpha=1$.

\subsection*{Convex duality and phase complementarity}

Let $\Psi_\alpha$ denote either the exact branch $\Phi_\alpha$ or the finite-support branch
$\widehat\Phi_\alpha$ and let $\Psi=\max_\alpha\Psi_\alpha$. For a prescribed mixture
density $\boldsymbol q$, the conjugate $\Psi^*(\boldsymbol
q)=\sup_{\boldsymbol\eta}\{\boldsymbol\eta^{\mathsf T}\boldsymbol q-\Psi(\boldsymbol\eta)\}$
gives the convex primal representation. At an attained interior optimum the Fenchel--Young
equality is equivalent to $\boldsymbol q\in\partial\Psi(\boldsymbol\eta)$. For a finite
collection of differentiable convex branches on a common open domain, the subdifferential
of the maximum is the convex hull of the active gradients\cite{rockafellar1970}. Consequently
the equilibrium conditions are
\begin{equation}
\boldsymbol q=\sum_\alpha S_\alpha\nabla\Psi_\alpha,\qquad
\sum_\alpha S_\alpha=1,\qquad
0\le S_\alpha\perp\Psi-\Psi_\alpha\ge0.
\label{eq:SIdualVolume}
\end{equation}
Here $a\perp b$ denotes $ab=0$. The active volume fractions determine the mixture density
state, while the shared dual coordinates impose thermal and chemical equilibrium.

For prescribed $T$, $P$ and overall mole fractions $\boldsymbol z$ ($\sum_i z_i=1$), fix
$\eta_0=-1/(RT)$ and denote the remaining coordinates by $\boldsymbol\eta_c$. The
pressure-constrained dual problem is
\begin{equation}
\max_{\boldsymbol\eta_c}\;\boldsymbol z^{\mathsf T}\boldsymbol\eta_c
\quad\text{subject to}\quad
\Psi_\alpha(\eta_0,\boldsymbol\eta_c)\le P/(RT)\quad\text{for all }\alpha.
\label{eq:SIdualTP}
\end{equation}
The objective is the reduced molar Gibbs energy at equilibrium. For the exact branches,
any feasible $\boldsymbol\eta_c$ gives a lower bound on the Gibbs energy of a phase mixture
through the branch Fenchel inequality: writing the Helmholtz-energy density as
$f_\alpha=u_\alpha-Ts_\alpha$, one has $f_\alpha/(RT)\ge\boldsymbol\eta_c^{\mathsf
T}\boldsymbol c_\alpha-\Psi_\alpha$. The bound follows by adding $P/(RT)$, multiplying by
the phase volume per total mole $v_\alpha$ and summing subject to $\sum_\alpha v_\alpha\boldsymbol
c_\alpha=\boldsymbol z$. Equality holds for conjugate states at the prescribed pressure
in every active phase.

At a solution in the interior of the dual domain, with differentiable finite branches and
a constraint qualification such as Slater feasibility, the Karush--Kuhn--Tucker conditions
of the convex problem are
\begin{equation}
\boldsymbol z=\sum_\alpha v_\alpha\nabla_{\boldsymbol\eta_c}\Psi_\alpha,
\qquad 0\le v_\alpha\perp P/(RT)-\Psi_\alpha\ge0.
\label{eq:SIdualKKT}
\end{equation}
The multipliers $v_\alpha$ have units of m$^3$\,mol$^{-1}$. Writing
$c_\alpha=\sum_i\partial_{\eta_i}\Psi_\alpha$, the molar phase fractions are $v_\alpha
c_\alpha$, and the volume fractions are $S_\alpha=v_\alpha/\sum_\gamma v_\gamma$. Thus
Eqs.~\eqref{eq:SIdualVolume} and \eqref{eq:SIdualKKT} give the common equilibrium closure
for density and pressure--composition specifications, respectively. These conditions
characterize global optima of the represented convex problem. The phase family and conjugate
domains must contain the equilibrium states of the reference model. The finite-support
error bounds in Supplementary Note~2 quantify fidelity to these reference branches. In a
conservation-law discretization the common dual coordinates and phase fractions enter the
global unknowns directly.

\section*{Supplementary Note 2:\\Finite-support approximation and error propagation}

We establish how support spacing, curvature and smoothing determine the approximation error
of a finite-support branch, and how this error propagates to the generated state, phase
classification and coexistence compositions.

\subsection*{Scaled branch representation}

We express the bounds in dimensionless coordinates. Throughout this note, $\widehat\varphi_M$
denotes the normalized smooth envelope before the shared offset is added. In physical units
the absolute potential error after addition of the offset is at most the unshifted bound
plus $|\Delta\Phi|$. The offset cancels from phase differences and its derivatives vanish. Let

\begin{equation}
\boldsymbol X=D_\eta^{-1}(\boldsymbol\eta-\boldsymbol\eta_{\rm ref}),\qquad
\varphi(\boldsymbol X)=\frac{\Phi(\boldsymbol\eta)-\Phi_{\rm ref}}{\Phi_s},
\label{eq:SIscaled}
\end{equation}
where $\boldsymbol\eta_{\rm ref}$ and $\Phi_{\rm ref}$ are the reference coordinate and potential,
$D_\eta$ is a diagonal coordinate-scaling matrix and $\Phi_s>0$ is the potential scale. The physical
state is recovered from the scaled gradient as
\begin{equation}
\boldsymbol q=\Phi_sD_\eta^{-\mathsf T}\nabla_X\varphi.
\label{eq:SIbackgrad}
\end{equation}
For a support at $\boldsymbol X_m$, we define $\varphi_m=\varphi(\boldsymbol X_m)$ and $\boldsymbol
g_m=\nabla\varphi(\boldsymbol X_m)$. The corresponding exact tangent support is
\begin{equation}
\ell_m(\boldsymbol X)=\varphi_m+\boldsymbol g_m^{\mathsf T}(\boldsymbol X-\boldsymbol X_m).
\end{equation}
For contact tangents of the represented convex branch, convexity gives
$\ell_m(\boldsymbol X)\le\varphi(\boldsymbol X)$ throughout its convexity domain. Consequently,
both the hard tangent envelope and its uncalibrated normalized log-sum-exp approximation
bound the exact
branch potential from below. The gap between the exact branch and a support plane is the Bregman
divergence\cite{bregman1967}:
\begin{equation}
D_\varphi(\boldsymbol X,\boldsymbol X_m)
=\varphi(\boldsymbol X)-\varphi(\boldsymbol X_m)-\nabla\varphi(\boldsymbol X_m)^{\mathsf
T}(\boldsymbol X-\boldsymbol X_m)
=\varphi(\boldsymbol X)-\ell_m(\boldsymbol X).
\label{eq:SIbregman}
\end{equation}
The hard finite-support branch is $H_M(\boldsymbol X)=\max_m\ell_m(\boldsymbol X)$, and the
normalized MAGE branch is
\begin{equation}
\widehat\varphi_M(\boldsymbol X)
=\bar\tau\log\left[\frac{1}{M}\sum_{m=1}^{M}
\exp\left(\frac{\ell_m(\boldsymbol X)}{\bar\tau}\right)\right],
\qquad \bar\tau=\tau/\Phi_s.
\label{eq:SIsmoothscaled}
\end{equation}
The standard log-sum-exp smoothing inequality\cite{nesterov2005,calafiore2020} gives
\begin{equation}
H_M-\bar\tau\log M\le\widehat\varphi_M\le H_M.
\label{eq:SIlsebound}
\end{equation}
The deficit of the smoothed branch relative to the hard envelope is
\begin{equation}
H_M-\widehat\varphi_M=\bar\tau\log\frac{M}{N_{\rm eff}},\qquad
N_{\rm eff}=\sum_{m=1}^{M}\exp\left(\frac{\ell_m-H_M}{\bar\tau}\right)\in[1,M],
\label{eq:SIdeficit}
\end{equation}
where $N_{\rm eff}$ is an effective count of the supports whose planes lie within approximately
$\bar\tau$ of the dominant plane. The smoothing deficit therefore depends on both the total support
count and the local support density; the bound $\bar\tau\log M$ corresponds to $N_{\rm eff}=1$. For
two competing branches, unequal deficits near coexistence shift the location of their intersection.

\subsection*{Potential error from support spacing and curvature}

We first bound the error in the potential itself. Let $\Omega\subset\mathbb R^d$ be the compact
scaled state domain of one physical phase. On a region containing the line segments from the
evaluation points to their associated supports, we assume that $\varphi$ is twice continuously
differentiable and that
\begin{equation}
0\preceq\nabla^2\varphi\preceq \Lambda I
\label{eq:SILsmooth}
\end{equation}
Let the support fill distance be
\begin{equation}
\delta_M=\sup_{\boldsymbol X\in\Omega}\min_m\|\boldsymbol X-\boldsymbol X_m\|_2.
\label{eq:SIfill}
\end{equation}
Taylor's theorem together with Eq.~\eqref{eq:SILsmooth} gives, for the nearest support,
\begin{equation}
0\le D_\varphi(\boldsymbol X,\boldsymbol X_m)\le\frac{\Lambda}{2}\|\boldsymbol X-\boldsymbol
X_m\|_2^2.
\end{equation}
Consequently,
\begin{equation}
0\le \varphi-H_M\le \frac{\Lambda}{2}\delta_M^2,
\label{eq:SIharderr}
\end{equation}
and combining Eqs.~\eqref{eq:SIlsebound} and \eqref{eq:SIharderr} yields the uniform MAGE potential
bound
\begin{equation}
0\le \varphi-\widehat\varphi_M
\le \underbrace{\frac{\Lambda}{2}\delta_M^2}_{\text{finite support}}
+\underbrace{\bar\tau\log M}_{\text{smoothing}}.
\label{eq:SIpotbound}
\end{equation}
For a branch with bounded curvature, the finite-support contribution is therefore second
order in the scaled support spacing. In physical units, after applying the common additive
pressure calibration,
\begin{equation}
|\Phi-\widehat\Phi|\le \Phi_s\varepsilon_\varphi+|\Delta\Phi|,
\qquad
|p-\widehat p|\le RT(\Phi_s\varepsilon_\varphi+|\Delta\Phi|)
\label{eq:SIpressurebound}
\end{equation}
at fixed temperature, where $\varepsilon_\varphi$ denotes the right-hand side of
Eq.~\eqref{eq:SIpotbound}.

\subsection*{Gradient and thermodynamic-state error}

Pressure follows from the value of the potential, whereas the density state that enters the material
balance follows from its gradient. To bound the gradient error quantitatively, we additionally
impose the nondegeneracy condition
\begin{equation}
\sigma I\preceq\nabla^2\varphi(\boldsymbol X)\preceq \Lambda I,
\qquad \sigma>0,
\label{eq:SIstrong}
\end{equation}
on a region containing every line segment between the query and a retained support. This
condition controls every contribution to the weighted gradient, including those from
supports with small weights. The MAGE weights satisfy
\begin{equation}
\pi_m\propto\exp\left[-\frac{D_\varphi(\boldsymbol X,\boldsymbol X_m)}{\bar\tau}\right],
\label{eq:SIgibbsweights}
\end{equation}
because $\ell_m=\varphi-D_\varphi$. If $D_* = \min_mD_\varphi(\boldsymbol X,\boldsymbol X_m)$, the
Gibbs variational inequality associated with Eq.~\eqref{eq:SIgibbsweights} gives
\begin{equation}
\sum_m\pi_mD_\varphi(\boldsymbol X,\boldsymbol X_m)
\le D_*+\bar\tau\log M
\le\frac{\Lambda}{2}\delta_M^2+\bar\tau\log M.
\label{eq:SIexpectedD}
\end{equation}
Strong convexity implies $D_\varphi\ge \sigma\|\boldsymbol X-\boldsymbol X_m\|^2/2$. The
$\Lambda$-smoothness condition gives
 $\|\nabla\varphi(\boldsymbol X_m)-\nabla\varphi(\boldsymbol X)\|\le \Lambda\|\boldsymbol
X_m-\boldsymbol X\|$. Since
$\nabla\widehat\varphi=\sum_m\pi_m\nabla\varphi(\boldsymbol X_m)$,
\begin{align}
\|\nabla\widehat\varphi-\nabla\varphi\|
&\le \Lambda\sum_m\pi_m\|\boldsymbol X_m-\boldsymbol X\|\\
&\le \Lambda\sqrt{\sum_m\pi_m\|\boldsymbol X_m-\boldsymbol X\|^2}\\
&\le \Lambda\sqrt{\frac{\Lambda\delta_M^2+2\bar\tau\log M}{\sigma}}.
\label{eq:SIgradbound}
\end{align}
Equation~\eqref{eq:SIgradbound} shows how support spacing, the upper and lower curvature
bounds and smoothing jointly control the gradient error. Through Eq.~\eqref{eq:SIbackgrad},
the corresponding
physical-state error is
\begin{equation}
\|\widehat{\boldsymbol q}-\boldsymbol q\|
\le \Phi_s\|D_\eta^{-\mathsf T}\|
 \Lambda\sqrt{\frac{\Lambda\delta_M^2+2\bar\tau\log M}{\sigma}}.
\label{eq:SIqbound}
\end{equation}
Decreasing $\sigma$ weakens the gradient-error bound and hence the control of state recovery
in directions of weak local curvature.

\subsection*{Validation thresholds and domain-wide bounds}

The bounds above apply over a compact domain under the stated curvature and support-coverage
assumptions, whereas greedy support construction is assessed on finite candidate and validation
sets. Let $\mathcal V$ be a validation $h_V$-net of $\Omega$,
\begin{equation}
h_V=\sup_{\boldsymbol X\in\Omega}\min_{\boldsymbol X_v\in\mathcal V}\|\boldsymbol X-\boldsymbol
X_v\|.
\end{equation}
If $\|\nabla\varphi\|\le G$ on a convex region containing $\Omega$ and all support states,
then both $\varphi$ and its hard tangent envelope are
$G$-Lipschitz; hence their difference is at most $2G$-Lipschitz. If the maximum hard-envelope gap on
validation states is $\theta_\Phi$, then
\begin{equation}
\sup_{\Omega}(\varphi-H_M)
\le \theta_\Phi+2Gh_V,
\label{eq:SIvalidationpot}
\end{equation}
and the smoothed branch satisfies
\begin{equation}
\sup_{\Omega}|\varphi-\widehat\varphi_M|
\le \theta_\Phi+2Gh_V+\bar\tau\log M.
\label{eq:SIvalidationpotsmooth}
\end{equation}
Once the validation fill distance is specified, this relation extends a finite validation
threshold to a bound over the full domain.

The analogous gradient bound follows from the covariance form of the Hessian. Let
\begin{equation}
D_g=\max_{m,n}\|\boldsymbol g_m-\boldsymbol g_n\|.
\end{equation}
The covariance form of the MAGE Hessian implies
\begin{equation}
\|\nabla^2\widehat\varphi\|\le\frac{D_g^2}{4\bar\tau},
\end{equation}
because the variance of any scalar projection of the weighted support slopes is at most one quarter
of the squared range of that projection. Therefore, if the maximum gradient error on $\mathcal V$ is
$\theta_g$,
\begin{equation}
\sup_{\Omega}\|\nabla\widehat\varphi-\nabla\varphi\|
\le \theta_g+\left(\Lambda+\frac{D_g^2}{4\bar\tau}\right)h_V.
\label{eq:SIvalidationgrad}
\end{equation}
Equations~\eqref{eq:SIvalidationpotsmooth} and \eqref{eq:SIvalidationgrad} give a direct
interpretation of validation thresholds: the threshold measures the resolved error on the validation
net, while $h_V$ accounts for unsampled gaps between validation states.

\subsection*{Potential-gap control during support selection}

At a fixed smoothing scale, let $w_{\alpha m j}$ be the normalized library weight of tangent
$m$ at probe $j$. Let $n_\alpha$ and $M_\alpha$ be the library and subset sizes, and let
$a_{\alpha j}=\sum_{m\in\mathcal S_\alpha}w_{\alpha m j}$. The exact change in the smoothed
branch when restricting it to the subset is
\begin{equation}
\delta_{\alpha j}=\tau_\alpha\log\!\left(\frac{a_{\alpha j}}{M_\alpha/n_\alpha}\right).
\end{equation}
Accordingly $d_j=\delta_{1j}-\delta_{2j}$ is the change in the phase margin. Adding one
candidate of library weight $w$ changes $\delta_{\alpha j}$ by $\tau_\alpha[\log(1+w/a_{\alpha
j})-\log(1+1/M_\alpha)]$. This identity supplies the candidate ranking in Supplementary
Note~3 and accounts for both the local tangent weight and the branch normalization. At
fixed scales, if $\Delta_{\mathcal S}$ and $\Delta_{\mathcal L}$ denote the selected and
complete-library phase margins in physical potential units, and $\Delta_{\rm ref}$ denotes
the exact margin, then
\begin{equation}
\Delta_{\mathcal S}-\Delta_{\rm ref}=(\Delta_{\mathcal S}-\Delta_{\mathcal L})+(\Delta_{\mathcal
L}-\Delta_{\rm ref}).
\label{eq:SIselectionDecomp}
\end{equation}
The first term is the selection drift; the second is the reference approximation error of
the complete library. The selection criteria control the former, whereas the support
coverage and smoothing bounds control the latter. The same decomposition applies at a fixed
tuned scale with its corresponding complete library.

\subsection*{Phase-margin criterion}

The potential difference between two competing branches determines phase classification.
For two physical phases $\alpha$ and $\beta$, we define the exact
and approximate branch differences
\begin{equation}
\Delta(\boldsymbol X)=\varphi_\alpha(\boldsymbol X)-\varphi_\beta(\boldsymbol X),\qquad
\widehat\Delta(\boldsymbol X)=\widehat\varphi_\alpha(\boldsymbol
X)-\widehat\varphi_\beta(\boldsymbol X).
\end{equation}
For the uncalibrated normalized envelopes, which are lower bounds of their exact convex
branches on the common scaled domain, write $e_\gamma=\varphi_\gamma-\widehat\varphi_\gamma^{(0)}$
with $0\le e_\gamma\le\varepsilon_\gamma$ for each branch $\gamma$. The common additive
calibration cancels from $\widehat\Delta$. Since $\widehat\Delta-\Delta=e_\beta-e_\alpha$,
the one-sided bound is
\begin{equation}
|\widehat\Delta-\Delta|\le\max\{\varepsilon_\alpha,\varepsilon_\beta\}\equiv\varepsilon_\Delta.
\label{eq:SImarginerr}
\end{equation}
Consequently,
\begin{equation}
|\Delta(\boldsymbol X)|>\varepsilon_\Delta
\quad\Longrightarrow\quad
\operatorname{sign}\widehat\Delta=\operatorname{sign}\Delta.
\label{eq:SIphasecertificate}
\end{equation}
Equation~\eqref{eq:SIphasecertificate} is a sufficient condition for correct phase
classification: away from
coexistence, no approximate branch crossing can occur where the exact phase margin $|\Delta|$
exceeds the
one-sided error bound $\varepsilon_\Delta$ on the branch difference.

\subsection*{Coexistence-location and phase-composition error}

To extend the phase-margin analysis to the coexistence point itself, consider a smooth
one-dimensional path $\boldsymbol X(s)$ crossing the coexistence manifold transversely,
for example along a chemical-coordinate direction at fixed temperature. Let $s^*$ satisfy
$\Delta(\boldsymbol X(s^*))=0$ and let $\widehat s$ satisfy $\widehat\Delta(\boldsymbol X(\widehat
s))=0$. Suppose
\begin{equation}
\left|\frac{\mathrm{d}}{\mathrm{d} s}\Delta(\boldsymbol X(s))\right|\ge\kappa>0
\label{eq:SIkappa}
\end{equation}
in the interval between $s^*$ and $\widehat s$. By Eq.~\eqref{eq:SImarginerr} and the mean-value
theorem,
\begin{equation}
|\widehat s-s^*|\le\frac{\varepsilon_\Delta}{\kappa}.
\label{eq:SIrootshift}
\end{equation}
The branch-crossing slope is the projection of the difference between the scaled density
states of the two phases onto the path direction:
\begin{equation}
\frac{\mathrm{d}\Delta}{\mathrm{d} s}
=\left(\nabla\varphi_\alpha-\nabla\varphi_\beta\right)^{\mathsf T}
\frac{\mathrm{d}\boldsymbol X}{\mathrm{d} s}.
\label{eq:SIcrossslope}
\end{equation}
Predictions of the coexistence location therefore become increasingly sensitive to potential
errors as the coexisting phase states approach one another, as expected near a critical point.

Let $V_X$ bound $\|\mathrm{d}\boldsymbol X/\mathrm{d} s\|$, let $\Lambda_\alpha$ be the curvature
bound of Eq.~\eqref{eq:SILsmooth} for branch $\alpha$ and let $\varepsilon_{g,\alpha}$ be a uniform
bound on the scaled-gradient approximation error of branch $\alpha$. The state error evaluated at
the displaced approximate coexistence point satisfies
\begin{equation}
\|\nabla\widehat\varphi_\alpha(\boldsymbol X(\widehat s))-
\nabla\varphi_\alpha(\boldsymbol X(s^*))\|
\le \varepsilon_{g,\alpha}+\Lambda_\alpha V_X\frac{\varepsilon_\Delta}{\kappa}.
\label{eq:SIcoexgrad}
\end{equation}
After the physical transformation in Eq.~\eqref{eq:SIbackgrad}, let $E_{c,\alpha}$ be a 1-norm bound
on the component-density error and let $c_{T,\alpha}=\sum_i c_i^\alpha$. Provided
$E_{c,\alpha}<c_{T,\alpha}$, the phase mole fractions satisfy
\begin{equation}
\|\widehat{\boldsymbol x}_\alpha-\boldsymbol x_\alpha\|_1
\le \frac{2E_{c,\alpha}}{c_{T,\alpha}-E_{c,\alpha}}.
\label{eq:SIxbound}
\end{equation}
Combining Eqs.~\eqref{eq:SIpotbound}, \eqref{eq:SIgradbound}, \eqref{eq:SIrootshift},
\eqref{eq:SIcoexgrad} and \eqref{eq:SIxbound} gives the following error chain from the support
parameters to the phase-composition error:
\begin{equation}
(\delta_M,\Lambda,\bar\tau,M,\sigma)
\longrightarrow
(\varepsilon_\Phi,\varepsilon_q)
\longrightarrow
\frac{\varepsilon_\Delta}{\kappa}
\longrightarrow
\varepsilon_{\rm phase\ composition}.
\label{eq:SIerrorchain}
\end{equation}
Support spacing and upper curvature control the local tangent approximation, and smoothing
adds a bounded error relative to the hard maximum. Lower curvature controls the gradient-error
bound. The phase margin governs stable-phase classification through
Eq.~\eqref{eq:SIphasecertificate}, and the branch-crossing slope $\kappa$ governs the
amplification of potential error into coexistence and composition errors. The bounds also
explain the sensitivity near criticality: as phase branches coalesce, the
branch curvature can increase, $\kappa$ approaches zero and $\sigma$ can become small.
Supplementary Table~\ref{tab:SIerrorbudget} summarizes these quantities.

\begin{table}[t]
\centering
\caption{\textbf{Quantities controlling the finite-support error budget.}}\label{tab:SIerrorbudget}
\small
\begin{tabularx}{\textwidth}{@{}p{0.16\textwidth}p{0.24\textwidth}X@{}}
\toprule
Quantity & Mathematical role & Consequence for MAGE accuracy\\
\midrule
$\delta_M$ & support fill distance & potential support error scales as $O(\Lambda\delta_M^2)$\\
$\Lambda$ & upper branch curvature & larger values increase both the tangent-approximation error and
the state-gradient error bound\\
$\bar\tau$ & dimensionless smoothing & contributes at most $\bar\tau\log M$ to the potential error
and broadens the distribution of support weights\\
$\sigma$ & lower local curvature & controls the conditioning of state recovery from the gradient
through Eq.~\eqref{eq:SIgradbound}\\
$h_V$ & validation fill distance & converts finite validation thresholds into domain-wide bounds\\
$|\Delta|$ & phase margin & if larger than $\varepsilon_\Delta$, the stable phase is classified
correctly\\
$\kappa$ & branch-crossing slope & converts branch-potential error into coexistence-location error\\
$c_T$ & phase total molar density & converts component-density error into mole-fraction error\\
\bottomrule
\end{tabularx}
\end{table}

\subsection*{Prescribed-pressure coexistence and phase-fraction sensitivity}

For binary coexistence at fixed $T$ and $P$, use the two chemical coordinates and define
\begin{equation}
\boldsymbol F(\boldsymbol\eta_c)=
\begin{pmatrix}\Phi_l-P/(RT)\\\Phi_g-P/(RT)\end{pmatrix},\qquad
J_F=\begin{pmatrix}c_1^l&c_2^l\\c_1^g&c_2^g\end{pmatrix}.
\label{eq:SIfixedP}
\end{equation}
Let $\boldsymbol\eta_c^*$ be an exact root and $\widehat{\boldsymbol\eta}_c$ a nearby
approximate root. If $J_*=J_F(\boldsymbol\eta_c^*)$ has smallest singular value $\gamma>0$
and $\|J_F-J_*\|\le\gamma/2$ along the segment joining the roots, the integrated Jacobian
gives $\|\boldsymbol
F(\widehat{\boldsymbol\eta}_c)\|\ge(\gamma/2)\|\widehat{\boldsymbol\eta}_c-\boldsymbol\eta_c^*\|$.
Therefore
\begin{equation}
\|\widehat{\boldsymbol\eta}_c-\boldsymbol\eta_c^*\|\le\frac{2}{\gamma}\sup\|\widehat{\boldsymbol
F}-\boldsymbol F\|,
\label{eq:SIfixedPbound}
\end{equation}
where the supremum is over the local region and the absolute branch-potential errors include
the additive calibration. The first-order displacement is $-J_*^{-1}(\widehat{\boldsymbol
F}-\boldsymbol F)(\boldsymbol\eta_c^*)$ for a smooth small perturbation. In particular,
$\det J_*=c_l c_g(x_1^l-x_1^g)$ for binary compositions. Coalescing phase endpoints can
lead to ill-conditioning of the coexistence equations. For a regular multicomponent
equilibrium the complete active pressure and material-balance equations give the corresponding
estimate. In a ternary two-phase system at fixed $T,P,\boldsymbol z$, let $\boldsymbol
w=(\eta_1,\eta_2,\eta_3,\beta)^{\mathsf T}$, where $\beta$ is the vapour molar phase
fraction. With $\boldsymbol x$ and $\boldsymbol y$ the normalized liquid and vapour
concentration vectors, use
\begin{equation}
\boldsymbol F(\boldsymbol w)=\begin{pmatrix}
\Phi_l-P/(RT)\\ \Phi_g-P/(RT)\\
(1-\beta)x_1+\beta y_1-z_1\\ (1-\beta)x_2+\beta y_2-z_2
\end{pmatrix}.
\label{eq:SIterF}
\end{equation}
Normalization ensures that the third component balance follows from the first two. Composition
derivatives follow from the branch Hessian $H_\alpha=\nabla^2_{\eta_c}\Phi_\alpha$. For
$\boldsymbol x^\alpha=\boldsymbol c^\alpha/c_\alpha$ and $c_\alpha=\boldsymbol 1^{\mathsf
T}\boldsymbol c^\alpha>0$,
\begin{equation}
D_{\eta_c}\boldsymbol x^\alpha=\frac{(I-\boldsymbol x^\alpha\boldsymbol 1^{\mathsf
T})H_\alpha}{c_\alpha}.
\end{equation}
Consequently,
\begin{equation}
J_F=\begin{pmatrix}
(\boldsymbol c^l)^{\mathsf T}&0\\
(\boldsymbol c^g)^{\mathsf T}&0\\
(1-\beta)\nabla x_1^{\mathsf T}+\beta\nabla y_1^{\mathsf T}&y_1-x_1\\
(1-\beta)\nabla x_2^{\mathsf T}+\beta\nabla y_2^{\mathsf T}&y_2-x_2
\end{pmatrix}.
\label{eq:SIterJ}
\end{equation}
All derivatives in this expression hold $T$ fixed. To compare pressure and material-balance
sensitivities, define $W=\operatorname{diag}(RT/P,RT/P,1,1)$ and $\boldsymbol G=W\boldsymbol
F$. The same equations and derivatives hold for the MAGE branches. At a regular interior
two-phase root $\boldsymbol w^*$ with $0<\beta<1$, let $\gamma=\sigma_{\min}(J_G(\boldsymbol
w^*))>0$. For a nearby approximate root $\widehat{\boldsymbol w}$, if $\|J_G-J_G(\boldsymbol
w^*)\|\leq\gamma/2$ along the joining segment, integration gives
\begin{equation}
\|\widehat{\boldsymbol w}-\boldsymbol w^*\|\leq
\frac{2}{\gamma}\sup\|\widehat{\boldsymbol G}-\boldsymbol G\|.
\label{eq:SIterBound}
\end{equation}
The first two scaled residual errors are relative branch-pressure errors; the last two
combine the liquid and vapour composition errors weighted by their mole fractions in the
mixture. On a local domain with $0\leq\beta\leq1$, bounds $E_{p,l},E_{p,g}$ on the absolute
pressure errors and $E_x,E_y$ on the Euclidean composition errors give $\|\widehat{\boldsymbol
G}-\boldsymbol G\|\leq[(E_{p,l}/P)^2+(E_{p,g}/P)^2+\max(E_x,E_y)^2]^{1/2}$. The final
column of Eq.~\eqref{eq:SIterJ} contains the tie-line separation, whereas its chemical-coordinate
block captures the thermodynamic response of the endpoints. Equation~\eqref{eq:SIcoexgrad}
extends to a multivariate displacement using the corresponding bound on the scaled
chemical-coordinate displacement.

For a two-phase feed with fixed composition $\boldsymbol z=(1-\beta)\boldsymbol
x+\beta\boldsymbol y$, let $\delta\boldsymbol x=\widehat{\boldsymbol x}-\boldsymbol x$,
$\delta\boldsymbol y=\widehat{\boldsymbol y}-\boldsymbol y$ and $\delta\beta=\widehat\beta-\beta$.
Subtraction of the two material balances gives the exact identity
\begin{equation}
\delta\beta(\widehat{\boldsymbol y}-\widehat{\boldsymbol x})=-[(1-\beta)\delta\boldsymbol
x+\beta\delta\boldsymbol y].
\label{eq:SIbeta}
\end{equation}
Thus, for any norm and a nonzero predicted tie line, the absolute phase-fraction error is
bounded by $[(1-\beta)\|\delta\boldsymbol x\|+\beta\|\delta\boldsymbol y\|]/\|\widehat{\boldsymbol
y}-\widehat{\boldsymbol x}\|$. For a component with distinct phase endpoints, the scalar
form gives the sensitivity of the lever rule. The relation links the endpoint error to
partitioning accuracy and separates the endpoint displacement from its amplification by
the tie-line geometry. Supplementary Note~3 evaluates these contributions for the largest
vapour-fraction discrepancy in the GERG sample of main-text Fig.~4.

\subsection*{Convergence of finite-support representations}

Consider a sequence of contact-support sets on a compact scaled branch domain with fixed
scaling and a uniform curvature bound $\Lambda$. If
\begin{equation}
\delta_M\longrightarrow0,\qquad \bar\tau_M\log M\longrightarrow0,
\label{eq:SIconvergence}
\end{equation}
then Eq.~\eqref{eq:SIpotbound} gives $\|\widehat\varphi_M-\varphi\|_\infty\to0$. Under the
uniform positive lower curvature bound $\sigma$ on the support--query segments,
Eq.~\eqref{eq:SIgradbound} gives $\|\nabla\widehat\varphi_M-\nabla\varphi\|_\infty\to0$.
In physical units, the same conclusions hold for the density states and, when $\Delta\Phi_M\to0$,
for the calibrated potentials. A common offset cancels from phase differences, so phase-margin
convergence follows directly from the unshifted potential bounds. On a compact subset where
the stable branch has a positive minimum margin over every competitor, its identity is
recovered for all sufficiently large support sets by Eq.~\eqref{eq:SIphasecertificate}.

Along a transverse coexistence path with a simple interior crossing, continuity and uniform
convergence of the phase difference preserve the opposite signs at the endpoints of a
sufficiently small bracketing interval. Approximate crossings therefore exist in the
interval for sufficiently large support sets, and Eq.~\eqref{eq:SIrootshift} gives their
convergence to the exact crossing. For a regular prescribed-pressure equilibrium, consider
the complete scaled equilibrium residual, including material balances where required.
Uniform potential and gradient convergence imply uniform residual convergence in a
neighbourhood with positive total phase densities. For an interior regular root $\boldsymbol
w^*$ of the scaled residual $\boldsymbol G$, write $A=J_G(\boldsymbol w^*)$ and choose a
closed ball of radius $r$ about $\boldsymbol w^*$ on which $\|I-A^{-1}J_G\|\leq1/2$. Uniform
residual convergence gives $\sup\|A^{-1}(\widehat{\boldsymbol G}-\boldsymbol G)\|\leq r/2$
for sufficiently large support sets. The continuous map $\boldsymbol w\mapsto\boldsymbol
w-A^{-1}\widehat{\boldsymbol G}(\boldsymbol w)$ then maps the ball into itself by integration
of the exact Jacobian. Brouwer's fixed-point theorem gives a zero of the approximate
residual in the ball. Equations~\eqref{eq:SIfixedPbound} and \eqref{eq:SIterBound} then
give convergence of the equilibrium coordinates. For interior two-phase states the ball
can be chosen with $0<\beta<1$. The normalized state bounds and Eq.~\eqref{eq:SIbeta} yield
convergence of compositions and phase fractions for nonzero tie-line separation. For
sequences satisfying Eq.~\eqref{eq:SIconvergence} and the stated regularity and scaling
conditions, these results establish convergence of potential values, first derivatives and
regular equilibria, providing explicit accuracy targets for support coverage and smoothing.

\subsection*{Numerical bound along liquid--liquid coexistence}

At the reference pressure and 300~K, the symmetric model has analytic common-tangent
compositions of $x^{L_1}=0.0707201817$ and $x^{L_2}=0.9292798183$.

For the regular solution, chemical potentials along each analytic coexistence branch
translate together by a dimensionless increment $h$. The derivative of the potential along
this line is the total concentration $c$. From the Helmholtz energy, $p=p_0+K[(c/c_0)^2-1]/2$,
and $dc/dh=RTc_0^2/K$. The restricted potential is therefore quadratic with directional
curvature $\Lambda_h=249.434$~mol\,m$^{-3}$. For each support plane, we evaluated the
minimum difference between this quadratic and the affine plane over the full pressure
interval of 4.5--5.5~MPa. For line supports with covering radius $\delta_h$, the resulting
upper bound is $\Lambda_h\delta_h^2/2+\tau\log M$.

\begin{table}[!htbp]\centering\small
\caption{\textbf{Potential bound along symmetric coexistence.} The two branches are related
by component exchange. Values shown for L1 use 1,201 pressures. Bound and error are in
mol\,m$^{-3}$.}\label{tab:SIlleBounds}
\begin{tabular}{@{}lrrrr@{}}\toprule
Planes & $\delta_h$ & Smoothing bound & Total bound & Measured maximum\\
\midrule
2 & 0.0204 & 0.008309 & 0.06022 & 0.05191\\
4 & 0.0101 & 0.01662 & 0.02935 & 0.02083\\
8 & 0.004939 & 0.02493 & 0.02797 & 0.02031\\
16 & 0.002521 & 0.03324 & 0.03403 & 0.01643\\
32 & 0.0009278 & 0.04155 & 0.04165 & 0.01722\\
55 & 0.0009278 & 0.04804 & 0.04815 & 0.01706\\
276 & 0.0009278 & 0.06738 & 0.06748 & 0.0364\\
\bottomrule\end{tabular}\end{table}

\begin{figure}[!htbp]\centering
\includegraphics[width=0.94\linewidth]{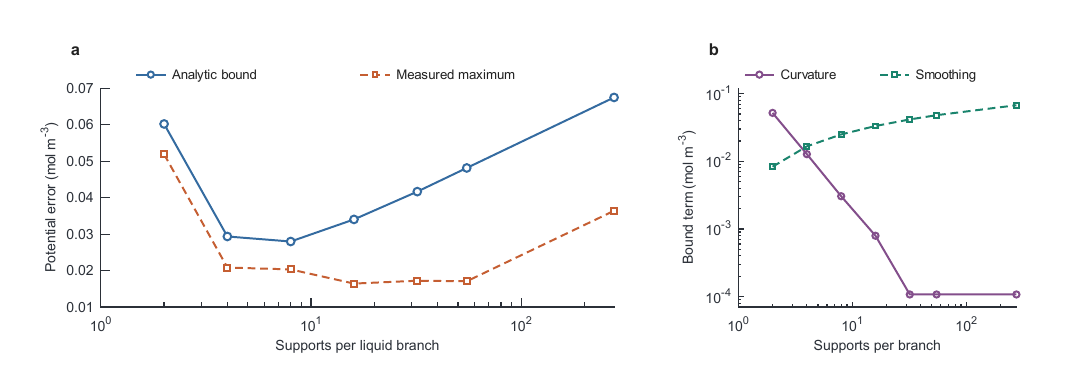}
\caption{\textbf{Potential error along the analytic coexistence path.} \textbf{a} Analytic
bound and maximum measured potential error over 1,201 pressures for the symmetric liquid
model. \textbf{b} Contributions from curvature and covering radius ($\Lambda\delta_M^2/2$)
and from smoothing ($\tau\log M$). Component exchange gives the same values for both branches.}
\label{fig:SIlleBounds}\end{figure}

The bound and the measured maximum over the support sequence are compared in Supplementary
Table~\ref{tab:SIlleBounds} and Supplementary Fig.~\ref{fig:SIlleBounds}. At the selected
size, the radius is $9.28\times10^{-4}$ and $\tau=0.011988$~mol\,m$^{-3}$. The bound is
0.06748~mol\,m$^{-3}$ and the maximum observed error on 1,201 pressures is 0.03640~mol\,m$^{-3}$.
These quantities characterize the potential error along the specified chemical-potential
path. The general gradient and phase-displacement results above retain their dependence
on the curvature and phase separation normal to coexistence.

\section*{Supplementary Note 3: Support construction, data partitions and equilibrium tests}

This note specifies the native tangent libraries and the selection rule shared by all
benchmarks, followed by the data partitions and equilibrium tests for each thermodynamic system.

\subsection*{Native tangent libraries and common selection rule}
Each candidate consists of its dual coordinates, native potential and gradient. The binary
PR library combines a regular grid of coexistence tangents with homogeneous states outside
the active smoothing width. The nonisothermal library varies temperature, pressure and
composition. The liquid--liquid library uses conjugate compositions of the symmetric
Helmholtz model. The brine library resolves the dilute-hydrogen regime and the aqueous
salt concentration. For GERG-2008, pressure--composition probes and saturation endpoints
determine the placement of native tangents; the competing branch pressure is evaluated at
each homogeneous inverse. The resulting states are evaluated with the reference EOS.

The PR liquid and vapour branches are distinguished by their low- and high-volume homogeneous
states; for a specified $T,p,\boldsymbol x$ these correspond to the smallest and largest
admissible compressibility roots when both are present (see Supplementary Note~6.4). The
reference coexistence endpoints identify the two branches on their common pressure range.
A mechanically regular, stable homogeneous root satisfies $(\partial p/\partial c)_{T,\boldsymbol
x}>0$; the second-order criterion for stability against component-density variations is
the Helmholtz curvature described in Supplementary Note~1. Branch-restricted contact follows
from the supporting inequality in that note. The symmetric liquid model uses the two
composition basins of the analytic free energy; the gas--brine model uses the aqueous
Gibbs-energy branch and the low-density hydrogen branch at the same thermodynamic reference
state. The GERG library distinguishes liquid and vapour through its homogeneous inverses
and saturation endpoints from the multiparameter reference EOS.

Within each branch of the selected GERG library, we evaluated every native plane at every
native coordinate. The maximum excess of the hard envelope over the native value was
$2.18\times10^{-11}$ and $1.77\times10^{-11}$~mol\,m$^{-3}$ for liquid and vapour,
respectively. The pairwise supporting-inequality check verifies contact within the finite
library to floating-point precision. The continuous-domain curvature estimates use the
branch regularity and contact conditions in Supplementary Notes~1 and 2. For the flow and
six-component benchmarks, initial sets of homogeneous thermodynamic states supply the two
candidate libraries. In the embedding control, selection uses the intrinsic chemical coordinates.

\begin{enumerate}
\item Scale coordinates by branchwise standard deviations, taking a unit scale on constant
coordinates. Rank the candidates by distance to the nearest calibration probe and assign
the closest 65\% to the near stratum. Use farthest-point sampling, allocating four of each
five choices to this stratum while both strata contain unused planes. Start the material
representations with 90\% of each library.
\item Compute the difference $d_j$ of the two branch changes derived in Supplementary
Note~2. Require the 99th percentile of $|d_j|$ to be at most the material tolerance
$\epsilon$ and its maximum to be at most $6\epsilon$. On the calibration states also compare
the number of physical failures and a continuous error with those of the complete library.
The failure count must be no greater than that of the complete library. For library error
$e$, the continuous error must be at most $\max(1.04e,e+0.0005)$.
\item If expansion is required, select the 48 probes with the largest $|d_j|$. For each
phase and candidate, let $g_j$ be the reduction in $|d_j|$ produced by its addition. Rank
candidates by $\langle g_j\rangle+0.25\max_j g_j+0.02\epsilon\max_j w_j$ on these probes.
Add up to 24 available tangents per branch. Subject to this cap, the batch size is one
quarter of the larger remaining branch pool, rounded upward. Repeat the checks.
\item Order retained tangents by their maximum library weight over the probes. Test deletions
of four tangents per branch and then one per branch, accepting a deletion only when both
checks are satisfied. Compare the smoothing multipliers 1, 0.999 and 1.001 on calibration
and separate tuning states. For each smoothing multiplier, require the failure count on
each partition to be at most the complete-library count on that partition. In GERG the
drift bounds on the shared geometry probes are checked again when the scale changes. Among
admissible choices, minimize the combined failure count, then the continuous tuning and
calibration errors in that order. If none of the three choices meets the criteria, retain
the full library at its input scale.
\end{enumerate}

Here $w_j$ is the candidate plane's normalized weight in its complete branch library at
probe $j$. The physical calibration check depends on the specified independent variables.
Binary PR and the regular solution compare phase classifications on either side of each
endpoint and measure endpoint displacement. The test feeds lie on both sides of each
reference endpoint, at methane mole-fraction offsets of $\pm0.0004$ for binary PR and
$\pm0.0008$ for nonisothermal PR. The corresponding composition offset for the regular
solution is $\pm0.0001$. The continuous endpoint metric is the mean absolute displacement
divided by the offset. The brine check compares bubble-point displacement with the composition
interval corresponding to a trace gas fraction. GERG and ternary PR flow use pressure--composition
equilibrium, the six-component test uses the pressure ordering at fixed chemical potentials,
and embedding uses the potential difference in intrinsic coordinates. GERG includes
additional shared geometry probes in the potential-gap check. The final evaluation applies
these physical checks to the retained model with fixed parameters.

\begin{table}[!htbp]\centering\small
\caption{\textbf{Input libraries and selected branch parameters.} Counts are per branch.
$\tau_\alpha$, $\Delta\Phi$ and the drift tolerance $\epsilon$ are in mol\,m$^{-3}$. A
single entry denotes equal smoothing scales. Paired entries give the values for the two
branches in the order listed.}\label{tab:SIdata}
\begin{tabular}{@{}lrrrrr@{}}\toprule
System & Library & Selected & $\tau_\alpha$ & $\Delta\Phi$ & $\epsilon$\\
\midrule
Binary PR & 1843/1840 & 1653/1651 & 0.001998 & 0.0135771 & 0.018\\
Nonisothermal PR & 2236/2110 & 2236/2110 & 1.4 & 9.64848 & 0.035\\
Liquid--liquid & 313/313 & 276/276 & 0.011988 & 0 & 0.003\\
Hydrogen--brine & 2940/1080 & 2940/1080 & 160/25 & 112.511 & 0.014\\
Ternary PR flow & 1283/1283 & 1149/1149 & 0.636363 & 3.80562 & 0.018\\
GERG-2008 & 2507/2493 & 2502/2488 & 1.70318/1.71834 & 19.2215 & 0.0001\\
Six-component PR & 71/71 & 60/60 & 0.031968 & 0 & 0.055\\
Embedding & 95/95 & 83/83 & 0.23976 & 0 & 0.06\\
\bottomrule\end{tabular}\end{table}

\subsection*{Binary coexistence and support-count tests}
For the fixed-temperature binary vapour--liquid equilibrium (VLE), the prescribed $T$ and $p$ fix
$\eta_0$ and $\Phi=p/(RT)$. The chemical coordinates are obtained by requiring both the liquid and
vapour MAGE branches to take this value, and the phase compositions follow from the branch
gradients. The same procedure is applied at each temperature of the nonisothermal benchmark.
We report relative branch errors on the independent homogeneous test states and phase-equilibrium
errors as absolute mole-fraction differences from the reference coexistence curve.

The 45 reference pressures in main-text Fig.~2a range from 0.55 to 4.95~MPa in steps of
0.10~MPa. Across the two selected branches, the minimum difference between a reference
pressure and a native support pressure is 70.5~Pa. The liquid and vapour endpoint MAEs at
these pressures are $4.11\times10^{-6}$ and $8.87\times10^{-6}$, and the maximum active-pressure
residual is $3.07\times10^{-8}$~Pa.

For the selected 3,304-support model, tests on 4,096 independent homogeneous states per
phase give mean relative potential errors of $3.57\times10^{-4}$ and $7.84\times10^{-3}$
for liquid and vapour, respectively. The corresponding mean relative component-density-vector
errors are $1.14\times10^{-3}$ and $2.49\times10^{-2}$ (main-text Fig.~2b,c).

\begin{figure}[!htbp]\centering
\includegraphics[width=\linewidth]{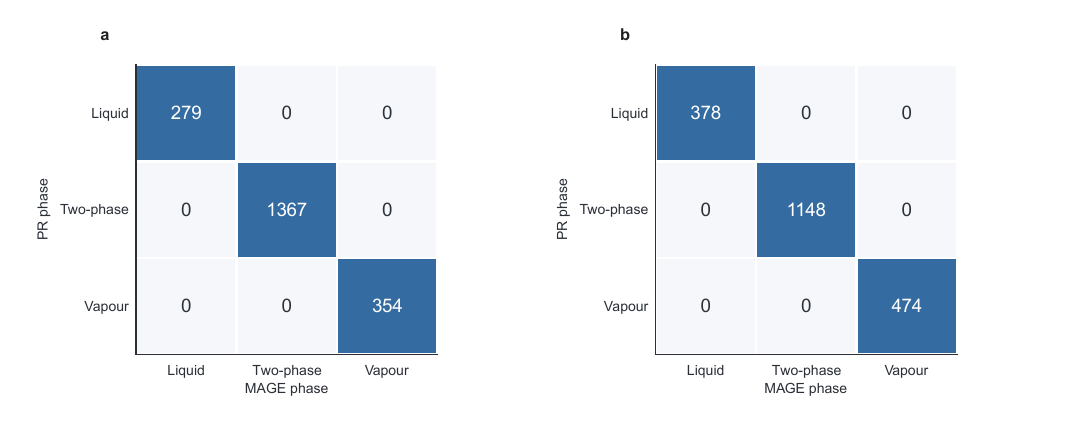}
\caption{\textbf{Phase-state agreement for methane + $n$-butane feeds.} \textbf{a} Isothermal
feeds at 300~K. \textbf{b} Nonisothermal feeds at 295--350~K. Rows give the PR phase and
columns the MAGE phase. Each entry counts individual feeds. Blue cells mark agreement.}
\label{fig:SIfeedPhase}\end{figure}

\begin{figure}[!htbp]\centering
\includegraphics[width=\linewidth]{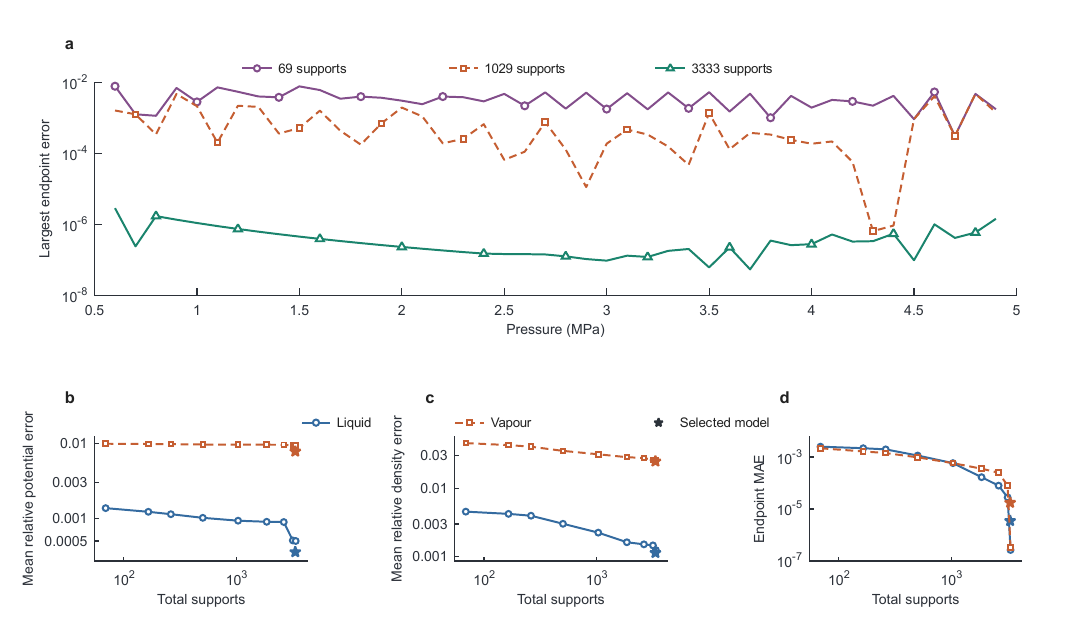}
\caption{\textbf{Support count and binary coexistence.} \textbf{a} Larger of the liquid
and vapour absolute methane mole-fraction errors at each of 44 interleaved pressures, for
three nested support counts. \textbf{b,c} Mean relative potential and component-density
errors on the independent homogeneous states. \textbf{d} Mean absolute error of each
endpoint over the 44 pressures. Lines in \textbf{b--d} join nested support sets. Filled
stars mark the selected 3,304-support model.}
\label{fig:SIsupportCount}
\end{figure}

The 44 midpoint pressures are $0.60,0.70,\ldots,4.90$~MPa and are distinct from the retained
support pressures on each branch. Both active MAGE branch pressures match the prescribed
pressure within $10^{-3}$~Pa at these 44 midpoints. The independent isothermal feed sample
comprises 279 liquid, 1,367 two-phase and 354 vapour states, with complete agreement between
PR and MAGE (Supplementary Fig.~\ref{fig:SIfeedPhase}a). For the support-count study, the
initialization fraction was 0.019. At each point along the expansion path, we used the
same 44 midpoint pressures and independent homogeneous branch tests. Supplementary
Fig.~\ref{fig:SIsupportCount} resolves the pressure dependence of the endpoint errors and
compares the branch and equilibrium errors along the support sequence, and Supplementary
Table~\ref{tab:SIbinaryMid} lists the endpoint errors, smoothing-deficit differences and
branch-pressure gaps at the midpoints.

\begin{table}[!htbp]\centering\small
\caption{\textbf{Binary coexistence at the 44 pressure midpoints.} The last row uses the
selected representation. Deficit differences are in mol\,m$^{-3}$ and pressure gaps are
evaluated at the reference dual states.}\label{tab:SIbinaryMid}
\begin{tabular}{@{}lrrrrr@{}}\toprule
Total & Liquid/vapour & $x$ MAE & $y$ MAE & Mean $|\Delta{\rm deficit}|$ & Gap (Pa)\\
\midrule
69 & 35/34 & 0.00249 & 0.00215 & $5.8\times10^{-5}$ & $1.28\times10^{3}$\\
165 & 83/82 & 0.00217 & 0.00163 & $9.56\times10^{-5}$ & 938\\
261 & 131/130 & 0.00196 & 0.00141 & 0.00021 & 697\\
501 & 251/250 & 0.00113 & 0.000965 & 0.000479 & 205\\
1029 & 515/514 & 0.000579 & 0.000577 & 0.000594 & 48.8\\
1845 & 923/922 & 0.000166 & 0.000353 & 0.000545 & 17.2\\
2613 & 1307/1306 & $7.88\times10^{-5}$ & 0.000247 & 0.000381 & 4.96\\
3141 & 1571/1570 & $2.78\times10^{-5}$ & $7.88\times10^{-5}$ & 0.000419 & 3.06\\
3333 & 1667/1666 & $2.64\times10^{-7}$ & $3.25\times10^{-7}$ & 0.00032 & 0.89\\
3304 & 1653/1651 & $3.49\times10^{-6}$ & $1.75\times10^{-5}$ & 0.000359 & 1.05\\
\bottomrule\end{tabular}\end{table}

The smoothing deficit of each branch is the hard envelope minus the normalized smooth
envelope at the reference coexistence coordinates, evaluated in the same gauge. For the
selected branches the mean deficits are 0.013486 and 0.013226~mol\,m$^{-3}$. The mean
absolute difference is 0.000359~mol\,m$^{-3}$. At the same coordinates the mean absolute
difference between the two smooth branch pressures is 1.05~Pa.

\subsection*{Nonisothermal ideal-gas caloric closure}

For methane + $n$-butane, the total Helmholtz-energy density is the sum of the Peng--Robinson (PR)
residual term\cite{peng1976} and the concentration-based ideal-gas contribution
\begin{equation}
 f^{\mathrm{id}}(T,\boldsymbol c)=RT\sum_i
c_i\left[\ln\!\left(\frac{c_iRT}{p^\circ}\right)-1\right]+\sum_i c_i g_i^\circ(T),\qquad
 g_i^\circ(T)=h_i^\circ(T)-Ts_i^\circ(T).
 \label{eq:SIidealHelmholtz}
\end{equation}
The ideal-gas heat capacities of the reference equations of state (EOS) follow the Planck--Einstein
form
\begin{equation}
 \frac{c_{p,i}^\circ(T)}{R}=1+a_i+\sum_k
n_{ik}\frac{(\theta_{ik}/T)^2\exp(-\theta_{ik}/T)}{[1-\exp(-\theta_{ik}/T)]^2}.
 \label{eq:SIPlanckEinstein}
\end{equation}
For methane\cite{setzmann1991} and $n$-butane\cite{buecker2006}, respectively, the parameters are
\begin{align}
 a_{\mathrm{CH_4}} &=3.0016, &\boldsymbol
n_{\mathrm{CH_4}}&=(0.008449,4.6942,3.4865,1.6572,1.4115),\\
 &&\boldsymbol\theta_{\mathrm{CH_4}}&=(648,1957,3895,5705,15080)\ \mathrm{K},\\
 a_{\mathrm{nC_4}} &=3.24680487, &\boldsymbol
n_{\mathrm{nC_4}}&=(5.54913289,11.4648996,7.59987584,9.66033239),\\
 &&\boldsymbol\theta_{\mathrm{nC_4}}&=(329.4040,1420.1737,2113.0894,4240.8573)\ \mathrm{K}.
\end{align}
These coefficients reproduce the ideal-gas contributions of the cited pure-fluid reference equations
of state over 295--350~K. All thermodynamic states are evaluated from the combined ideal-gas and
residual Helmholtz model with these analytic correlations.

The reference state is $T^\circ=298.15$~K and $p^\circ=1.0\times10^5$~Pa. Each ideal-gas reference
function is normalized such that $h_i^\circ(T^\circ)=0$~J mol$^{-1}$ and
$s_i^\circ(T^\circ,p^\circ)=0$~J mol$^{-1}$ K$^{-1}$, so that
\begin{equation}
 h_i^\circ(T)=\int_{T^\circ}^{T}c_{p,i}^\circ(T')\,\mathrm{d} T',\qquad
 s_i^\circ(T)=\int_{T^\circ}^{T}\frac{c_{p,i}^\circ(T')}{T'}\,\mathrm{d} T'.
 \label{eq:SIcaloricIntegrals}
\end{equation}
The integrals are evaluated analytically. Defining $x_{ik}=\theta_{ik}/T$,
$x_{ik}^\circ=\theta_{ik}/T^\circ$, $B(x)=1/(\exp x-1)$ and $F(x)=x/(\exp x-1)-\ln(1-\exp(-x))$
gives
\begin{align}
 h_i^\circ(T)&=R(1+a_i)(T-T^\circ)+R\sum_k n_{ik}\theta_{ik}[B(x_{ik})-B(x_{ik}^\circ)],\\
 s_i^\circ(T)&=R(1+a_i)\ln(T/T^\circ)+R\sum_k n_{ik}[F(x_{ik})-F(x_{ik}^\circ)].
\end{align}
Setting $h_i^\circ=s_i^\circ=0$ at the reference state defines sensible-energy and entropy
origins common to both phases. This convention preserves the PR pressure and the vapour--liquid
coexistence conditions.

\subsection*{Nonisothermal equilibrium and homogeneous derivatives}
The evaluation grid of 222 temperatures and 82 pressures per temperature contains 18,204
states spanning 295--350~K and 1--7~MPa. The equilibrium calculation recovered ordered
endpoints at every state, with both active-phase pressures matching the prescribed pressure
within $10^{-3}$~Pa. All 222 evaluation temperatures are distinct from the retained support
temperatures on both branches. In the temperature--pressure plane, all evaluation points
lie inside the convex hull of the support points of each branch. To compute the distances
in Supplementary Fig.~\ref{fig:SInonisoDistance}, we scale the reference coexistence dual
coordinates by the standard deviations of the corresponding branch supports. The median
and 95th-percentile distances are 0.0341 and 0.0657, respectively, for liquid, and 0.0719
and 0.1805 for vapour. An independent sample of 2,000 feeds contains 378 liquid, 1,148
two-phase and 474 vapour states. MAGE and PR agree on all three phase labels throughout
this sample (Supplementary Fig.~\ref{fig:SIfeedPhase}b). Supplementary Table~\ref{tab:SInonisoVLE}
summarizes the endpoint and homogeneous-branch errors.

\begin{figure}[!htbp]\centering
\includegraphics[width=0.94\linewidth]{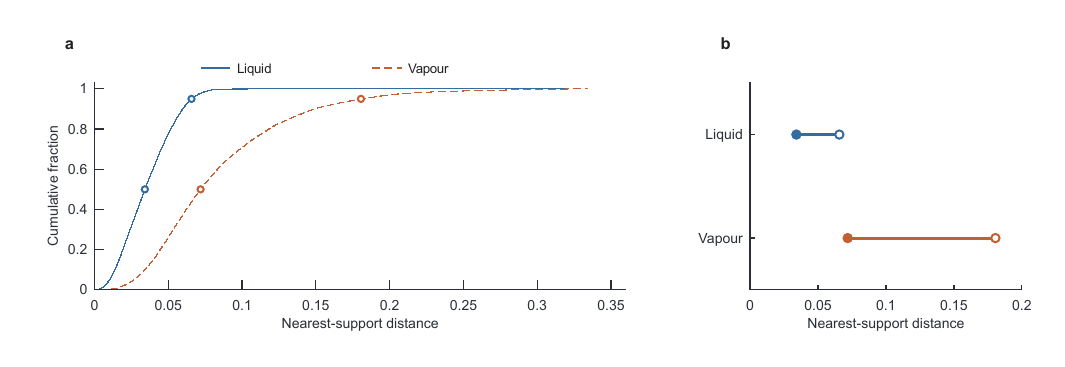}
\caption{\textbf{Support coverage of the nonisothermal coexistence grid.} \textbf{a}
Cumulative distributions of the nearest-support distance in scaled dual coordinates for
18,204 equilibrium states on each branch. Circles mark the median and 95th percentile.
\textbf{b} The same two quantiles on a common distance axis. Filled and open circles mark
the median and 95th percentile, respectively.}
\label{fig:SInonisoDistance}\end{figure}

\begin{table}[!htbp]\centering\small
\caption{\textbf{Nonisothermal coexistence and homogeneous-branch errors.} Endpoint errors
use absolute mole fractions. Energy errors are absolute.}\label{tab:SInonisoVLE}
\begin{tabular}{@{}lrrrr@{}}\toprule
Observable & States & Mean & 95th percentile & Maximum\\
\midrule
Liquid endpoint mole fraction & 18204 & $5.59\times10^{-5}$ & 0.000167 & 0.0014\\
Liquid relative potential & 2688 & 0.00206 & 0.00514 & 0.0342\\
Liquid relative density vector & 2688 & 0.00764 & 0.0172 & 0.0289\\
Liquid molar energy (kJ\,mol$^{-1}$) & 2688 & 0.0698 & 0.277 & 0.535\\
Vapour endpoint mole fraction & 18204 & $7.53\times10^{-5}$ & 0.000196 & 0.00436\\
Vapour relative potential & 2688 & 0.00421 & 0.0118 & 0.024\\
Vapour relative density vector & 2688 & 0.035 & 0.0934 & 0.166\\
Vapour molar energy (kJ\,mol$^{-1}$) & 2688 & 0.198 & 0.539 & 1.06\\
\bottomrule\end{tabular}\end{table}

\subsection*{Phase classification on independently sampled feeds}
Each binary feed specifies the overall methane mole fraction $z$. For each model, the
coexisting liquid and vapour methane mole fractions $x$ and $y$ delimit the two-phase
interval. A feed with $z<x$ is liquid, a feed with $z>y$ is vapour and a feed between the
endpoints contains both phases. Each sample comprises 2,000 independent draws with $z$
uniform on $[0,1]$. In the isothermal sample $T=300$~K and $p$ is uniform over 0.55--4.95~MPa.
In the nonisothermal sample $T$ and $p$ are drawn independently and uniformly over 295--350~K
and 1--7~MPa. For each feed, we compared the PR and MAGE labels, checked the ordering of
the endpoints and evaluated each active-branch pressure residual against the $10^{-3}$~Pa
tolerance. All pairs of endpoints were ordered. The largest pressure residuals were
$4.52\times10^{-8}$~Pa at fixed temperature and $4.27\times10^{-7}$~Pa at variable temperature.

\subsection*{Branch scales and the shared pressure offset}
The tangent sum is evaluated with physical potentials and gradients. Each material's
smoothing scale and the shared additive offset are reported in Supplementary Table~\ref{tab:SIdata}.
Standard deviations used for distance calculations affect the selection metric. A coordinate
transformation preserves the tangent value when the gradient is transformed by the inverse
transpose. In the flow library the shared offset is the mean difference between the candidate
reference potentials and their smooth library estimates across both phases. At 320~K, this
offset shifts both phase pressures by 10.13~kPa and contributes to their saturation-weighted
mixture pressure. Its common value cancels in the pressure difference controlling phase
competition at fixed dual coordinates. At a prescribed pressure it enters the absolute
branch pressure constraints and hence the solution coordinates, as in Eq.~\eqref{eq:SIfixedP}.
Supplementary Table~\ref{tab:SIflowBranch} reports the density-gradient errors of the two
flow branches on independent homogeneous states.

\begin{table}[!htbp]\centering\small
\caption{\textbf{Density gradients of the flow branches.} Each branch is evaluated at 4,096
independent homogeneous states. Component columns give absolute MAEs in
mol\,m$^{-3}$.}\label{tab:SIflowBranch}
\begin{tabular}{@{}lrrrr@{}}\toprule
Branch & Relative vector mean & CH$_4$ & CO$_2$ & $n$-C$_4$\\
\midrule
Liquid & 0.00218 & 6.83 & 8.48 & 13.7\\
Vapour & 0.0141 & 5.04 & 3.69 & 2.59\\
\bottomrule\end{tabular}\end{table}

\subsection*{GERG-2008 phase-equilibrium evaluation}
The reference is the GERG-2008 fundamental Helmholtz model\cite{kunz2012}, evaluated for
CO$_2$ + CH$_4$ + $n$-butane with Thermopack\cite{wilhelmsen2017}. At 320~K the feed
pressure is uniform over 1.5--10~MPa. The unnormalized CO$_2$, CH$_4$ and $n$-butane
fractions are drawn independently and uniformly over $[0.35,0.65]$, $[0.10,0.30]$ and
$[0.15,0.35]$, then divided by their sum. The 4,990-support representation and evaluation
protocol were fixed before the confirmatory test. The 5,000 independent feeds were generated
with the prescribed seed 32017605419 and evaluated in full (main-text Fig.~4a,b). MAGE and
GERG agree on the number of equilibrium phases for all 5,000 feeds, comprising 1,037
single-phase and 3,963 two-phase states. The phase-number agreement and partitioning
statistics are based on this complete sample.

On 20,000 independent homogeneous states per branch, the mean relative potential errors
are $7.41\times10^{-4}$ (liquid) and $3.42\times10^{-4}$ (vapour); the corresponding mean
relative component-density-vector errors are 0.00888 and 0.0117 (main-text Fig.~4c).

For a vapour volume fraction $S_g$, the overall composition follows by normalizing
$(1-S_g)\boldsymbol c_l+S_g\boldsymbol c_g$. For a two-phase solution, the vapour molar
phase fraction is $S_g c_g/[(1-S_g)c_l+S_g c_g]$, where $c_\alpha=\sum_i c^\alpha_i$. An
independent evaluation of the native tangent sum checks the maximum composition residual
($10^{-10}$ tolerance), the mixture and active-phase pressure residuals and the excess of
either branch pressure over the specified pressure ($10^{-3}$~Pa tolerance for each). Phase
volumes are nonnegative. Coalescing branches with both maximum composition separation and
relative total-density separation below $10^{-4}$ are treated as a single phase.

\begin{table}[!htbp]\centering\small
\caption{\textbf{GERG phase compositions.} Mean and maximum absolute mole-fraction errors
on the 3,963 two-phase feeds.}\label{tab:SIgergClass}
\begin{tabular}{@{}lrrrr@{}}\toprule
Component & Liquid mean & Liquid max. & Vapour mean & Vapour max.\\
\midrule
CO$_2$ & 0.00256 & 0.168 & 0.000866 & 0.0192\\
CH$_4$ & 0.00123 & 0.0380 & 0.000805 & 0.00975\\
$n$-C$_4$ & 0.00279 & 0.206 & 0.000670 & 0.0236\\
\bottomrule\end{tabular}\end{table}

\begin{figure}[!htbp]\centering
\includegraphics[width=0.94\linewidth]{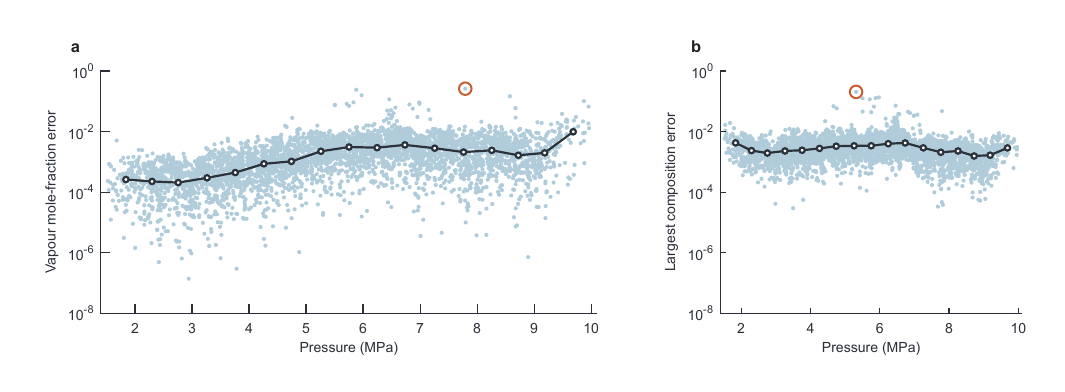}
\caption{\textbf{Partitioning on the GERG two-phase feeds.} \textbf{a,b} Absolute errors
in vapour molar phase fraction and in the least accurately predicted phase composition,
respectively, for each of 3,963 feeds. The composition error is the largest of six
componentwise errors across the two phases. Lines connect the medians within 0.5~MPa
intervals; rings mark the maxima.}
\label{fig:SIgergSplit}\end{figure}

All 5,000 feeds satisfy the closure tests. Independent evaluation of the native tangents
gives a maximum overall composition residual of $5.00\times10^{-11}$ and a maximum
active-phase pressure residual of $4.98\times10^{-4}$~Pa. Among the 3,963 two-phase feeds,
the median, mean and 95th percentile of the absolute vapour-fraction error are 0.00127,
0.003031 and 0.00898, respectively; the maximum is 0.26256. Supplementary
Table~\ref{tab:SIgergClass} gives the corresponding componentwise errors for both phases,
and Supplementary Fig.~\ref{fig:SIgergSplit} shows how the errors vary with pressure.

At 7.79232~MPa, the liquid and vapour mole fractions of $n$-butane are $(0.25415,0.12747)$
for GERG and $(0.20658,0.12443)$ for MAGE (Supplementary Table~\ref{tab:SIgergTie}). Both
intervals contain the feed mole fraction, 0.18018. The lever rule $\beta=(z-x)/(y-x)$ gives
vapour molar phase fractions of 0.58392 and 0.32136, respectively. The largest separation
in any component is 0.12668 for GERG and 0.08214 for MAGE. Of the 3,963 MAGE two-phase
feeds, 41 (including this feed) have a maximum component separation this small or smaller,
corresponding to approximately 1.03\% of the sample.

At the GERG coexistence coordinates, the MAGE liquid and vapour pressures differ from the
prescribed pressure by $-36.0$ and $-0.545$~kPa. To separate the local gradient error from
the shift of the coexistence coordinates, write $\boldsymbol x$ for the normalized branch
gradient and use the exact decomposition
\begin{equation}
\widehat{\boldsymbol x}(\widehat{\boldsymbol\eta})-\boldsymbol x(\boldsymbol\eta^*)=
[\widehat{\boldsymbol x}(\boldsymbol\eta^*)-\boldsymbol x(\boldsymbol\eta^*)]+[\widehat{\boldsymbol
x}(\widehat{\boldsymbol\eta})-\widehat{\boldsymbol x}(\boldsymbol\eta^*)].
\label{eq:SIgergDecomp}
\end{equation}
Evaluation of the retained tangents at the GERG coordinates gives a liquid $n$-butane mole
fraction of 0.25924 and a vapour mole fraction of 0.12478. The local liquid endpoint error
is therefore $+0.00509$, while the coordinate displacement contributes $-0.05267$, producing
the net error $-0.04758$. The corresponding vapour contributions are $-0.00269$ and $-0.00035$.

At the reference coordinates, the liquid-minus-vapour pressure-error difference of about
$-35.5$~kPa perturbs the phase-equality condition, whereas the individual branch-pressure
errors perturb the prescribed-pressure constraints in Eq.~\eqref{eq:SIterF}. In the coupled
pressure and composition-balance equations, these perturbations displace the common
coexistence coordinates, predominantly shifting the liquid endpoint and shortening the
predicted tie line. Equation~\eqref{eq:SIbeta} then gives a vapour-fraction change of
approximately $-0.263$ from these endpoint errors. 

At the MAGE intersection, the overall composition and mixture-pressure residuals are
$1.39\times10^{-16}$ and $2.79\times10^{-9}$~Pa. The scaled MAGE Jacobian of Eq.~\eqref{eq:SIterJ},
with the first two rows divided by $P/(RT)$ and the remaining rows unscaled, has singular
values $(2.948,0.454,0.0773,0.0257)$ at the GERG coordinates and reference vapour fraction,
and $(2.904,0.194,0.0528,0.0161)$ at the MAGE solution. The corresponding 2-norm condition
numbers are 115 and 180. The Euclidean norm of the scaled MAGE residual at these points
is $1.16\times10^{-2}$ and $3.34\times10^{-16}$, respectively. Centred differences with a
dimensionless increment of $10^{-6}$ agree with the analytic Jacobians to a maximum entrywise
difference of $2.65\times10^{-8}$. The two evaluations of the MAGE Jacobian characterize
the local sensitivity of the represented equilibrium equations at the reference and predicted
states. Together with the endpoint-error decomposition, they distinguish the accuracy of
numerical closure from the fidelity of the thermodynamic representation relative to GERG.

\begin{table}[!htbp]\centering\small
\caption{\textbf{Tie lines at the largest vapour-fraction discrepancy.} Overall and phase
mole fractions at 320~K and 7.79232~MPa.}\label{tab:SIgergTie}
\begin{tabular}{@{}lrrrrr@{}}\toprule
Component & Feed & GERG liquid & MAGE liquid & GERG vapour & MAGE vapour\\
\midrule
CO$_2$ & 0.70629 & 0.66191 & 0.69108 & 0.73791 & 0.73841\\
CH$_4$ & 0.11353 & 0.08393 & 0.10235 & 0.13463 & 0.13716\\
$n$-C$_4$ & 0.18018 & 0.25415 & 0.20658 & 0.12747 & 0.12443\\
\bottomrule\end{tabular}\end{table}

\section*{Supplementary Note 4: Fully implicit ternary Darcy formulation}

We specify the discrete formulation of the coupled ternary benchmark and the independent
fully implicit PR reference, together with the quantitative criteria used to compare the
two solutions.

\subsection*{Physical problem and assumptions}

The benchmark considers a CH$_4$ + CO$_2$ + $n$-butane mixture, described by the PR EOS, at 320~K in
a one-dimensional domain of length $L_x=0.6$~m. The domain contains $N_g=6$ equal cell-centred
finite volumes of width $\Delta x=0.1$~m. The porosity and absolute permeability are constant,
$\phi=0.20$ and $K=10^{-15}$ m$^2$. Gravity, capillary pressure and molecular diffusion are omitted
so that the benchmark isolates compressible accumulation, pressure-driven advection, thermodynamic
phase competition and phase appearance and disappearance. The inlet and outlet pressures are fixed
at 3.2~MPa and 2.0~MPa, respectively. The initial pressure is the linear cell-centre profile
\begin{equation}
P_j^0=P_{\rm in}+\frac{x_j}{L_x}(P_{\rm out}-P_{\rm in}),
\qquad x_j=(j-\tfrac12)\Delta x.
\label{eq:SIinitialP}
\end{equation}
The initial overall composition is
\begin{equation}
\boldsymbol z^0=(0.04,0.08,0.88).
\end{equation}
The three inlet-composition stages are
\begin{align}
0<t\le 12{,}000~{\rm s}:&\quad \boldsymbol z_{\rm in}=(0.15,0.15,0.70),\\
12{,}000<t\le 30{,}000~{\rm s}:&\quad \boldsymbol z_{\rm in}=(0.04,0.08,0.88),\\
30{,}000<t\le 60{,}000~{\rm s}:&\quad \boldsymbol z_{\rm in}=(0.65,0.20,0.15).
\end{align}
The phase viscosities and relative permeabilities are
\begin{equation}
\zeta_l=2.5\times10^{-4}\ {\rm Pa\,s},\qquad
\zeta_g=1.5\times10^{-5}\ {\rm Pa\,s},
\end{equation}
\begin{equation}
k_{rl}=(1-S_g)^2,\qquad k_{rg}=S_g^2.
\label{eq:SIrelperm}
\end{equation}
For the 12,000~s first injection, both independent global solvers use the same fixed sequence of 92
backward-Euler time steps. The time increments are specified for each of the three injection
stages A, B and C.
Stage A uses
\begin{equation}
500,\,375,\,6\times506.25,\,683.4375,\,9\times750,\,654.0625\ {\rm s};
\end{equation}
Stage B uses
\begin{equation}
3\times500,\,4\times675,\,13\times750,\,5\times562.5,\,750,\,487.5\ {\rm s};
\end{equation}
and Stage C uses
\begin{equation}
500,\,8\times375,\,506.25,\,683.4375,\,32\times750,\,2\times562.5,\,185.3125\ {\rm s}.
\end{equation}
The sequences contain 19, 27 and 46 steps, respectively, and sum exactly to the three stage
durations, yielding 92 common time levels for both solvers.

\subsection*{MAGE primary variables and thermodynamic closure}

At fixed temperature, the thermal Massieu coordinate $\eta_0=-1/(RT)$ is constant. In each cell $j$,
the MAGE primary-variable vector is
\begin{equation}
\boldsymbol y_j=
(\eta_{{\rm CH_4},j},\eta_{{\rm CO_2},j},\eta_{{\rm nC_4},j},S_{g,j},P_j)^{\mathsf T}.
\label{eq:SIprimary}
\end{equation}
There are therefore five unknowns per cell and $5N_g=30$ unknowns in the global nonlinear system.
The two physical phase branches use the same chemical Massieu coordinates. For $\alpha\in\{l,g\}$,
\begin{equation}
p_{\alpha,j}=RT\widehat\Phi_\alpha(\boldsymbol\eta_{c,j}),\qquad
c_{i,j}^{\alpha}=\frac{\partial\widehat\Phi_\alpha}{\partial\eta_i}(\boldsymbol\eta_{c,j}),
\label{eq:SIphaseprops}
\end{equation}
where $\boldsymbol\eta_c=(\eta_{\rm CH_4},\eta_{\rm CO_2},\eta_{\rm nC_4})$. Evaluating
both branches at a common $\boldsymbol\eta_c$ ensures equality of the candidate-phase
chemical potentials by construction. The component inventory
per unit bulk volume is
\begin{equation}
C_{i,j}=\phi\left[(1-S_{g,j})c_{i,j}^l+S_{g,j}c_{i,j}^g\right].
\label{eq:SIinventory}
\end{equation}

The physical pressure $P_j$ is an independent global unknown and satisfies the closure equation
\begin{equation}
R_{P,j}=P_j-\left[(1-S_{g,j})p_{l,j}+S_{g,j}p_{g,j}\right]=0.
\label{eq:SIpressureclosure}
\end{equation}
A single bounded complementarity equation in natural-map form determines the phase
state\cite{lauser2011}
\begin{equation}
R_{{\rm ph},j}=S_{g,j}-\Pi_{[0,1]}\left[
S_{g,j}+\frac{p_{g,j}-p_{l,j}}{P_{\rm scale}}
\right]=0,
\qquad P_{\rm scale}=1~{\rm MPa}.
\label{eq:SInaturalmap}
\end{equation}
Equation~\eqref{eq:SInaturalmap} is equivalent to
\begin{align}
S_g=0 &\quad\Longrightarrow\quad p_l\ge p_g,\label{eq:SIsingleL}\\
0<S_g<1 &\quad\Longrightarrow\quad p_l=p_g,\label{eq:SItwophase}\\
S_g=1 &\quad\Longrightarrow\quad p_g\ge p_l.\label{eq:SIsingleG}
\end{align}
The complementarity residual thus handles phase appearance and disappearance through a
single equation.
A cell moves continuously from the liquid boundary $S_g=0$ into the interior $0<S_g<1$ when the
branch-pressure margin reaches zero; it leaves the two-phase manifold when the natural-map solution
returns to either saturation bound. This algebraic equation applies in all phase regimes.

\subsection*{Darcy velocities and component fluxes}

For phase $\alpha$, we define the mobility
\begin{equation}
\lambda_\alpha(S_g)=\frac{k_{r\alpha}(S_g)}{\zeta_\alpha}.
\end{equation}
At an internal face $j+1/2$, the phase Darcy velocity is
\begin{equation}
u_{\alpha,j+1/2}
=-K\lambda_{\alpha,j+1/2}^{\rm up}
\frac{P_{j+1}-P_j}{\Delta x},
\label{eq:SIdarcyface}
\end{equation}
where the mobility is taken from the upstream cell for that phase. The component molar flux is
\begin{equation}
F_{i,j+1/2}=\sum_{\alpha\in\{l,g\}}
c_{i,j+1/2}^{\alpha,{\rm up}}u_{\alpha,j+1/2}.
\label{eq:SIcomponentflux}
\end{equation}
At the inlet face, the half-cell pressure gradient uses $P_{\rm in}$ and $P_1$, and inflowing phase
properties are evaluated from the prescribed stage composition after a one-time boundary-state
mapping. At the outlet face, the half-cell gradient uses $P_{N_g}$ and $P_{\rm out}$; for
outward flow, the phase properties are those of the last cell. Internal component
source terms are zero.

\subsection*{Fully implicit conservation residual}

Backward-Euler time integration and cell-centred finite volumes give, for component $i$ in cell $j$,
\begin{equation}
R_{i,j}^{n+1}=
C_{i,j}^{n+1}-C_{i,j}^{n}
+\frac{\Delta t_n}{\Delta x}
\left(F_{i,j+1/2}^{n+1}-F_{i,j-1/2}^{n+1}\right)=0.
\label{eq:SIconservation}
\end{equation}
For three components, Eqs.~\eqref{eq:SIconservation} provide three equations per cell. Together with
Eqs.~\eqref{eq:SIpressureclosure} and \eqref{eq:SInaturalmap}, the closed MAGE system is
\begin{equation}
\left\{
R_{{\rm CH_4},j},
R_{{\rm CO_2},j},
R_{{\rm nC_4},j},
R_{P,j},
R_{{\rm ph},j}
\right\}=\boldsymbol{0},
\qquad j=1,\ldots,6.
\label{eq:SIclosedsystem}
\end{equation}
In Eq.~\eqref{eq:SIclosedsystem}, direct evaluation of $p_\alpha$ and $\boldsymbol c^\alpha$
from the preconstructed MAGE branches replaces the nested nonlinear thermodynamic solve.

\subsection*{Analytic and generalized Jacobian}

The phase-branch derivatives used in the global Jacobian are available analytically. At fixed $T$,
\begin{equation}
\frac{\partial p_\alpha}{\partial\eta_k}=RTc_k^\alpha,
\qquad
\frac{\partial c_i^\alpha}{\partial\eta_k}
=H_{ik}^\alpha,
\qquad
H^\alpha=\nabla_{\eta_c}^2\widehat\Phi_\alpha\succeq0.
\label{eq:SIbranchjac}
\end{equation}
The inventory derivatives are
\begin{align}
\frac{\partial C_i}{\partial\eta_k}
&=\phi\left[(1-S_g)H_{ik}^l+S_gH_{ik}^g\right],\\
\frac{\partial C_i}{\partial S_g}
&=\phi(c_i^g-c_i^l).
\end{align}
The pressure-closure derivatives are
\begin{align}
\frac{\partial R_P}{\partial\eta_k}
&=-RT\left[(1-S_g)c_k^l+S_gc_k^g\right],\\
\frac{\partial R_P}{\partial S_g}&=p_l-p_g,\qquad
\frac{\partial R_P}{\partial P}=1.
\end{align}
Let
\begin{equation}
\omega=S_g+(p_g-p_l)/P_{\rm scale},
\end{equation}
and let $\chi$ denote a generalized derivative of $\Pi_{[0,1]}(\omega)$: $\chi=1$ for $0<\omega<1$,
$\chi=0$ outside the interval and $\chi=1/2$ at an endpoint within the numerical equality tolerance.
Then
\begin{align}
\frac{\partial R_{\rm ph}}{\partial S_g}&=1-\chi,\\
\frac{\partial R_{\rm ph}}{\partial\eta_k}
&=-\frac{\chi RT}{P_{\rm scale}}(c_k^g-c_k^l),\\
\frac{\partial R_{\rm ph}}{\partial P}&=0.
\end{align}
The upwind flux derivatives use the current upwind state in each nonlinear linearization.
We assemble the resulting block-local generalized Jacobian as a dense matrix and supply
it to a bound-constrained trust-region least-squares solver for the global residual. The
bounds used in every cell are
\begin{equation}
-0.45\le\eta_{\rm CH_4}\le3.35,\quad
-0.20\le\eta_{\rm CO_2}\le3.00,\quad
-0.45\le\eta_{\rm nC_4}\le1.55,
\end{equation}
\begin{equation}
0\le S_g\le1,\qquad
1.25\ {\rm MPa}\le P\le3.25\ {\rm MPa}.
\end{equation}
The variable scaling is $(1,1,1,0.2,10^6\,{\rm Pa})$ per cell. The nonlinear solver uses step,
cost and gradient tolerances of $10^{-11}$ and at most 5,000 function evaluations per time step.

Component residuals are scaled by 1,500~mol m$^{-3}$ and the pressure-closure residual by
$10^5$~Pa. The acceptance threshold for the maximum scaled residual is $10^{-6}$; the
largest observed MAGE value is $1.22\times10^{-7}$.

\subsection*{Thermodynamic evaluation in the coupled residual}

The offline construction provides a finite representation of the map for each phase
\begin{equation}
\boldsymbol\eta_c\longmapsto
\left(p_\alpha(\boldsymbol\eta_c),\boldsymbol
c^\alpha(\boldsymbol\eta_c),H^\alpha(\boldsymbol\eta_c)\right).
\end{equation}
During time integration, all thermodynamic quantities in Eqs.~\eqref{eq:SIclosedsystem}
and \eqref{eq:SIbranchjac} are evaluated directly using matrix-vector operations, exponentials,
normalized weights and reductions over the fixed support set. Common chemical coordinates
encode chemical equilibrium, the branch values determine pressure competition, and
Eq.~\eqref{eq:SInaturalmap} determines the phase amount globally. Before integration, a
PR flash at each of the six initial states and three inlet compositions supplies the
starting point for a four-variable MAGE inverse solve for $\boldsymbol\eta_c$ and $S_g$
at prescribed pressure and composition.

\subsection*{Independent fully implicit PR global reference}

The reference calculation solves the same conservation problem using a conventional
formulation with nested local equilibrium calculations. Each cell has three global unknowns,
\begin{equation}
\boldsymbol y_j^{\rm PR}=(P_j,\xi_{1,j},\xi_{2,j})^{\mathsf T},
\end{equation}
where unconstrained composition coordinates are mapped to the ternary simplex by
\begin{equation}
z_1=\frac{e^{\xi_1}}{1+e^{\xi_1}+e^{\xi_2}},\quad
z_2=\frac{e^{\xi_2}}{1+e^{\xi_1}+e^{\xi_2}},\quad
z_3=\frac{1}{1+e^{\xi_1}+e^{\xi_2}}.
\label{eq:SIsoftmaxz}
\end{equation}
For every state $(P,T,\boldsymbol z)$ required in a global residual evaluation, a local PR flash is
solved before the accumulation and flux terms are assembled. The local algorithm, built on analytic
PR cubic roots\cite{peng1976,michelsen1982a,michelsen1982b}, proceeds as follows. (i) The
equilibrium ratios are initialized with the Wilson correlation or with the $\ln K$ values stored at
the last accepted time step; during the global nonlinear iterations of a time step, this
initialization is held fixed. (ii) The Rachford--Rice equation $f(\beta)=0$ is solved for the vapour
molar phase fraction $\beta$: $\beta=0$ if $f(0)\le0$, $\beta=1$ if $f(1)\ge0$, and otherwise
$\beta$ is obtained by 30 bisection steps. (iii) The fugacity equations are solved by at most 35
successive-substitution iterations with a convergence threshold of $3\times10^{-9}$, and a damping
factor of 0.6 is applied while the error exceeds 0.15; if the threshold is not reached within 35
iterations, the current iterate is passed to the phase decision. (iv) For the overall composition
$\boldsymbol z$, the dimensionless Gibbs energy $g_s=\sum_i z_i(\ln z_i+\ln\phi_i)$ is evaluated for
every valid PR cubic root, and the root with the lowest $g_s$ is retained as the homogeneous
candidate. (v) The state is classified as two-phase only if $10^{-7}<\beta<1-10^{-7}$ and the
dimensionless Gibbs energy $g_2$ of the two-phase candidate satisfies $g_2<g_s-10^{-8}$; otherwise,
it is classified as single-phase with the lowest-Gibbs homogeneous root, labelled vapour if its
compressibility factor satisfies $Z_s>0.5$ and liquid otherwise. The phase decision therefore
compares the Gibbs energy of the two-phase candidate with those of the homogeneous PR roots
at the feed composition. For a two-phase flash with vapour molar phase
fraction $\beta$, liquid and vapour compositions $\boldsymbol x$ and $\boldsymbol y$, and total
phase molar densities $c_T^l$ and $c_T^g$, the volume-based gas saturation used in the Darcy
equations is
\begin{equation}
S_g=\frac{\beta/c_T^g}{(1-\beta)/c_T^l+\beta/c_T^g},
\label{eq:SIbetatoS}
\end{equation}
and
\begin{equation}
c_i^l=x_i c_T^l,\qquad c_i^g=y_i c_T^g.
\end{equation}
Single-phase states return $S_g=1$ (vapour) or $S_g=0$ (liquid) with the properties of the selected
homogeneous root. Local elimination of the phase variables by the flash leaves three
component conservation equations per cell in the global PR residual.

The same bound-constrained trust-region least-squares framework is used for the global reference
system. The unconstrained composition coordinates satisfy $-7\le\xi_i\le7$, and the pressure is
bounded between 1.3 and 3.3~MPa. These are solver-variable bounds; the physical inlet and outlet
pressures remain 3.2 and 2.0~MPa, respectively, and the MAGE variable bounds are specified above.
The global solver forms a dense two-point finite-difference Jacobian for the 18 PR unknowns,
with a relative difference step of $2\times10^{-5}$. Exact-state caching reuses the flashes
of cells whose local variables are fixed during a column perturbation. The step, cost and
gradient tolerances
($10^{-11}$) and the residual acceptance threshold ($10^{-6}$) are shared by both solvers,
with up to 5,000 function evaluations per step. The largest observed PR scaled residual
was $6.82\times10^{-9}$.

\paragraph{Verification of the accepted PR equilibrium states.}
For both injection schedules, we repeated the local flashes at the accepted cell pressures
and overall compositions in chronological order, using the same initialization and carrying
the accepted $\ln K$ values forward for each cell. The resulting phase states, vapour mole
fractions and gas saturations reproduced the corresponding trajectory values. For the
returned two-phase states, we evaluated the maximum log-fugacity residual
$r_f=\max_i|\ln(f_i^l/f_i^g)|$ using the final phase compositions and their PR fugacity
coefficients, and the component mole-balance residual $r_z=\max_i|z_i-[(1-\beta)x_i+\beta
y_i]|$. The latter quantifies local phase-partitioning closure, whereas the relative mass
imbalance in Supplementary Table~\ref{tab:SIflowresults} quantifies conservation over a flow step.

Phase stability was assessed through the dimensionless tangent-plane distance (TPD) at the
specified temperature and pressure. For a trial composition $\boldsymbol w$ and an occupied
reference phase of composition $\boldsymbol a$, the objective is
\begin{equation}
D(\boldsymbol w;\boldsymbol a)=\sum_i w_i\left[\ln(w_i\phi_i(\boldsymbol
w))-\ln(a_i\phi_i(\boldsymbol a))\right],
\label{eq:SItpdcheck}
\end{equation}
with zero-mole-fraction terms evaluated by their continuous limits. For a single-phase
state, the occupied phase supplies the reference; for a two-phase state, each occupied
phase supplies a separate reference. The search evaluates all admissible PR roots at 5,151
trial compositions spanning the ternary simplex, including its edges and vertices. Candidate
minima are then refined from multiple initial compositions along the liquid and vapour
root branches. Across both injection schedules, all 1,110 accepted PR cell--time states
met the local successive-substitution stopping criterion within 17 iterations. Across the
620 two-phase states, the maximum $r_f$ was $1.93\times10^{-9}$ and the maximum $r_z$ was
$1.09\times10^{-10}$. For all accepted states, every dimensionless TPD found by the
multistart search was at or above the verification threshold of $-10^{-8}$. The smallest
values of $D$ were reproduced at the composition of the other coexisting phase and were
comparable in magnitude to the fugacity residuals. Supplementary Table~\ref{tab:SIflowresults}
summarizes the checks for both schedules.

\subsection*{Phase-event definition and comparison}

Phase labels describe the converged saturation in each cell. They are defined as
\begin{equation}
S_g<10^{-4}:\ \text{liquid},\qquad
10^{-4}\le S_g\le1-10^{-4}:\ \text{two-phase},\qquad
S_g>1-10^{-4}:\ \text{vapour}.
\end{equation}
A phase event is a change in this classification between successive common time levels.
For the 12,000~s first injection, the two global solutions contain $92\times6=552$ common
cell--time states.
The labels agree at all 552 states, and all 24 phase events match in time step, cell and
transition type. The complete event list at the specified $10^{-4}$
threshold is given in Supplementary Table~\ref{tab:SIevents}.

In cell 1, the four transitions occur at accepted times of 500, 13,000, 30,500 and 48,189.7~s.

\begin{table}[!htbp]\centering\small
\caption{\textbf{Matched phase events with a 12,000~s first injection.} Entries give
accepted step and time (h) for each cell. L, LV and V denote liquid, two-phase and vapour
states.}\label{tab:SIevents}
\begin{tabular}{@{}lrrrr@{}}\toprule
Cell & L$\to$LV & LV$\to$L & L$\to$LV & LV$\to$V\\
\midrule
1 & 1 (0.139) & 21 (3.611) & 47 (8.472) & 75 (13.386)\\
2 & 4 (0.524) & 23 (3.938) & 49 (8.681) & 81 (14.636)\\
3 & 6 (0.806) & 25 (4.312) & 52 (8.993) & 85 (15.469)\\
4 & 8 (1.087) & 28 (4.917) & 54 (9.201) & 87 (15.886)\\
5 & 9 (1.277) & 33 (5.958) & 56 (9.446) & 88 (16.094)\\
6 & 10 (1.485) & 41 (7.521) & 57 (9.636) & 89 (16.303)\\
\bottomrule\end{tabular}\end{table}

\begin{table}[!htbp]\centering\small
\caption{\textbf{Ternary flow solutions and accepted PR equilibrium checks for two injection
schedules.} Pressure errors are relative. Mole-fraction and saturation errors are absolute.
Operation counts cover online integration. The PR flash counts L/LV/V refer to liquid,
two-phase and vapour states. Fugacity and component mole-balance residuals $r_f$ and $r_z$
are evaluated on returned two-phase solutions. The dimensionless TPD minima and counts
below $-10^{-8}$ refer to the multistart search in Eq.~\eqref{eq:SItpdcheck} using each
occupied phase as a reference.}\label{tab:SIflowresults}
\begin{tabular}{@{}lrr@{}}\toprule
Observable & 12,000~s first injection & 8,000~s first injection\\
\midrule
Accepted steps & 92 & 93\\
Cell--time states & 552 & 558\\
Matching labels & 552 & 558\\
Matching events & 24 & 24\\
Mean pressure error & 0.000635 & 0.000615\\
Mean component error & 0.00077 & 0.00075\\
Maximum component error & 0.0777 & 0.0807\\
Mean saturation error & 0.00152 & 0.00126\\
Maximum scaled residual, MAGE & $1.22\times10^{-7}$ & $1.47\times10^{-7}$\\
Maximum scaled residual, PR & $6.82\times10^{-9}$ & $7.43\times10^{-9}$\\
Maximum relative mass imbalance, MAGE & $4.9\times10^{-13}$ & $2.81\times10^{-13}$\\
Maximum relative mass imbalance, PR & $5.56\times10^{-9}$ & $5.56\times10^{-9}$\\
MAGE branch evaluations & 25828 & 27090\\
Executed PR flashes & 22113 & 22022\\
\midrule
\multicolumn{3}{@{}l}{\textit{Accepted PR equilibrium checks}}\\
PR flash phase counts (L/LV/V) & 176/323/53 & 208/297/53\\
States reaching 35 flash iterations & 0 & 0\\
Maximum flash iterations & 17 & 17\\
Maximum log-fugacity residual $r_f$ & $1.87\times10^{-9}$ & $1.93\times10^{-9}$\\
Maximum component mole-balance residual $r_z$ & $1.08\times10^{-10}$ & $1.09\times10^{-10}$\\
Minimum dimensionless TPD found & $-1.70\times10^{-9}$ & $-1.77\times10^{-9}$\\
States with TPD below $-10^{-8}$ & 0/552 & 0/558\\
\bottomrule\end{tabular}\end{table}

\begin{figure}[!htbp]\centering
\includegraphics[width=0.94\linewidth]{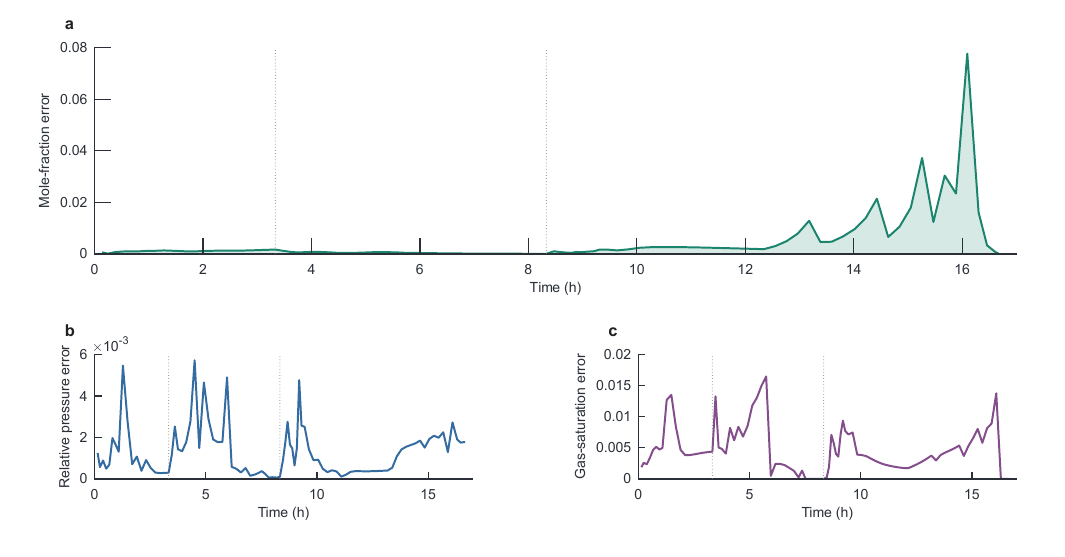}
\caption{\textbf{Cellwise errors during the three-stage injection.} \textbf{a} Maximum
absolute overall mole-fraction error across all six cells and three components. \textbf{b,c}
Maximum relative pressure and absolute gas-saturation errors across the cells. Vertical
lines mark the switches to liquid wash and vapour-rich injection.}
\label{fig:SIflowTime}\end{figure}

Supplementary Fig.~\ref{fig:SIflowTime} shows the maximum errors across all six cells as
functions of time. Pressure decreases from inlet to outlet at every accepted step. The
maximum overall-composition discrepancy of 0.0777 occurs for $n$-butane in cell 6 at
57,939.7~s, when both solutions contain two phases. Both solutions transition to vapour
at the next step, locating the discrepancy during passage of the final front at the outlet.

\subsection*{Nonlinear convergence and the smoothing scale}
\begin{figure}[!htbp]\centering
\includegraphics[width=\linewidth]{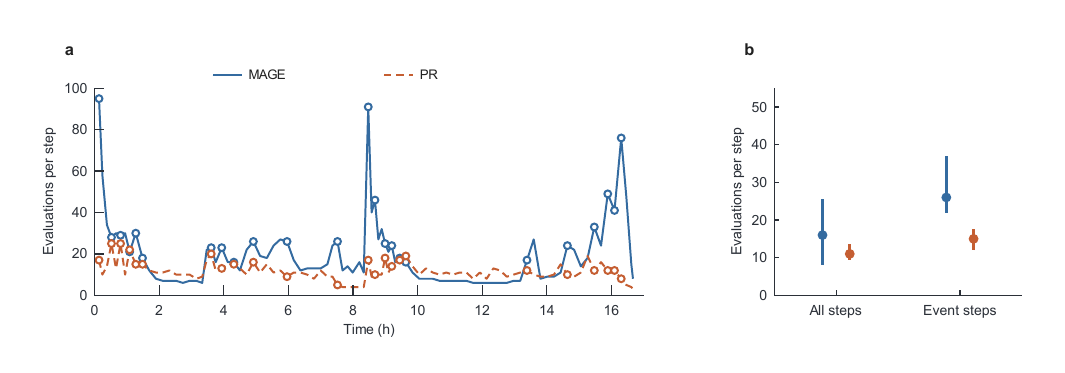}
\caption{\textbf{Nonlinear function evaluations per accepted step.} \textbf{a} MAGE and
PR function counts on the 92-step flow trajectory. Open circles mark steps with a phase
transition. \textbf{b} Medians (dots) and interquartile ranges (bars) over all steps and
over steps containing a transition. The optimizer reports function evaluations separately
from the finite-difference calls used to form the PR Jacobian.}
\label{fig:SIglobalSteps}
\end{figure}

We count nonlinear iterations as outer trust-region updates from the initial iterate. On
the 92-step trajectory MAGE required 1,615 such updates (median 14 per step) and PR required
768 (median 8). For steps containing a phase transition, the respective medians were 22.5
and 11. A trust-region update may require several residual evaluations. The median
solver-reported function counts per accepted step were 16 (MAGE) and 11 (PR), rising to
26 and 15, respectively, on steps with a phase event. Supplementary Fig.~\ref{fig:SIglobalSteps}
shows these counts for every accepted step. The time grid contains 92 accepted steps.
Summing the optimizer-reported function counts gives 1,856 for MAGE and 1,063 for PR. The
solver excludes the perturbation calls used to form numerical Jacobians from this count.
Direct counting gives 1,952 MAGE and 16,255 PR global residual evaluations. Each PR residual
requests one flash per cell, giving 97,530 requests. Of these requests, the cache serves
75,969, leaving 21,561 flashes executed within residual evaluations. Evaluating the six
accepted states after each of 92 steps adds 552 flashes for a total of 22,113. The PR EOS
is evaluated 439,265 times. MAGE makes 25,828 branch evaluations and zero nested flashes
during integration (main-text Fig.~5f). Each branch evaluation forms the potential and the
required derivatives from the supports. These operation counts distinguish the equilibrium
work performed by the two formulations. The two global systems also differ in structure:
MAGE has 30 global unknowns, including an explicit bounded phase variable in each cell,
whereas the PR system has 18 unknowns and resolves the phase split inside each residual evaluation.

The branch Hessian is the softmax-weighted covariance of the support gradients divided by
$\tau$. A smaller $\tau$ narrows the transition between dominant planes and concentrates
curvature where two planes have similar values. With the supports, shared offset, time
grid and solver tolerances fixed, we integrated the same trajectory at $\tau=0.636363$,
$0.5090904$ and $0.3818178$~mol\,m$^{-3}$. The solver accepted all 92 steps at each scale.
The total nonlinear iteration counts were 1,615, 1,805 and 1,960, with corresponding medians
of 14, 16 and 16. The optimizer used 1,856, 2,061 and 2,243 function evaluations, respectively.
The maximum scaled residuals over the 92 steps were $1.22\times10^{-7}$, $1.20\times10^{-7}$
and $1.21\times10^{-7}$, respectively. At the selected scale the mean absolute overall
composition and saturation errors relative to the PR trajectory were $7.70\times10^{-4}$
and $1.52\times10^{-3}$, and the mean absolute pressure error was 1.65~kPa. The other
scales increased the total iteration count by 11.8\% and 21.4\% while meeting the same
closure criterion. The smoothing bound relative to the hard envelope decreases with $\tau$
at fixed supports (Eq.~\eqref{eq:SIlsebound}), while the gradients and the branch intersections
depend on the distribution of the support weights. State accuracy and nonlinear iteration
counts therefore provide complementary criteria for selecting a smoothing scale.

\paragraph{Elapsed time and reuse.}
For each schedule, we timed four pairs of complete runs on the same Windows workstation
using one BLAS thread and reversing the solver order in successive pairs. Timing started
before solver construction and ended after the final solution and diagnostics had been
written, including boundary and initial-state mapping. The median elapsed times for MAGE
and PR were 4.58 and 20.97~s with first injections of 12,000~s, and 4.83 and 20.36~s with
first injections of 8,000~s. Each repetition reproduced the phase labels and all 24 phase
events. The PR-to-MAGE ratios of the median times were 4.58 and 4.22, respectively, for
these implementations.

First-use timings include PR candidate generation, support selection and the complete
trajectory with a 12,000~s first injection. With the complete candidate cloud, the calculation
took 21.51~s, comprising 10.87~s for reference-state sampling and evaluation, 0.93~s for
preselection and pressure calibration, 5.18~s for final support construction, and 4.53~s
for the trajectory. Because the input requires only the first 1,331 liquid and 1,572 vapour
candidates, sampling can stop when both prefixes are complete. This reduced the number of
reference draws from 2,417 to 1,776 and the total to 18.35~s, including 7.76~s for candidate
generation. Both sampling strategies produced identical thermodynamic libraries and selected
potentials. Both matched all 552 PR labels and all 24 events. Each first-use time is a
single end-to-end measurement. The paired trajectory timings quantify the cost of integrating
trajectories with reusable selected supports.

\section*{Supplementary Note 5:\\Additional thermodynamic and flow-trajectory evaluations}

\subsection*{5.1 Composite hydrogen--brine reference model}

We evaluated the finite-support construction for the H$_2$--H$_2$O--neutral NaCl gas--brine
system using a composite Gibbs-energy reference model. The reference model combines
Helmholtz-energy equation-of-state
(HEOS) contributions for water\cite{wagner2002} and hydrogen\cite{leachman2009}, a Pitzer activity
model for the aqueous phase\cite{pitzer1973,pitzer1984} and correlations for the Henry
constant\cite{fernandezprini2003}, the salting-out effect\cite{chabab2020} and the partial molar
volume\cite{brelvi1972} of hydrogen within a single extensive Gibbs-energy function. The
composite reference implementation evaluates each state at prescribed $T$, $p$ and liquid
salinity. The constituent thermodynamic terms provide the component chemical potentials,
from which the extensive Gibbs energy $G$ is formed. Phase-specific volume correlations
provide $V$ and convert component amounts to molar densities $c_i=n_i/V$. The internal-energy
density is evaluated as $[G-T(\partial G/\partial T)_{p,\boldsymbol n}]/V-p$, and the
native Massieu value is $\Phi=p/(RT)$. In online evaluation, the finite-support branches
generate pressure as $p_\alpha=RT\widehat\Phi_\alpha$, with gradients and Hessians obtained
from weighted moments of the stored support states.
The conserved salt variable is the neutral NaCl formula-unit concentration, and the gas phase is
described by a low-density hydrogen approximation; these definitions specify the thermodynamic model
represented in this benchmark.

Candidate and calibration states sample the reference system across temperature, pressure
and liquid salinity. The common selection rule uses a branch-gap tolerance of 0.014~mol\,m$^{-3}$
and tests the bubble-point displacement in the dilute-hydrogen limit.

The constituent thermodynamic models supply the support coefficients during offline
construction of the library. Online branch evaluation then uses these stored coefficients
to compute affine plane values, normalized exponential weights and weighted density moments.
For a branch with $M_\alpha$ supports and $d$ independent dual coordinates, evaluation of
the plane values, potential and gradient requires arithmetic work of order $M_\alpha d$;
evaluation of a dense Hessian by weighted covariance requires work of order $M_\alpha d^2$.
Thus, at fixed support count and coordinate dimension, the arithmetic cost per branch
evaluation is independent of the internal complexity of the reference thermodynamic model.
At a prescribed coordinate dimension, the thermodynamic evaluation work of an online
equilibrium solve therefore depends on the representation size and the number of branch evaluations.

The bubble-point calibration uses a fixed aqueous salt-to-water ratio. The selected
representation retains the full libraries of 2,940 liquid and 1,080 gas supports. The
candidate library, calibration and tuning states, and held-out test states occupy distinct
data partitions. The tie-line sample uses the held-out reference pool, including its tie
lines and homogeneous phase states, and is evaluated with the fixed support sets, smoothing
scales and shared offset. The independent confirmation sample comprises 512 tie lines
generated using random seed 314927, with temperature sampled uniformly from 295 to 423~K,
pressure log-uniformly from 1 to 45~MPa, and salinity uniformly from 0.05 to 5~mol\,kg$^{-1}$.
For each of these two samples, the overall mole fractions follow from $\boldsymbol
z=(1-\beta)\boldsymbol x+\beta\boldsymbol y$, where $\boldsymbol x$ and $\boldsymbol y$
are the liquid and vapour mole fractions. Each of the 512 tie lines defines seven feeds
at vapour fractions $10^{-5}$, 0.01, 0.1, 0.5, 0.9, 0.99 and 0.99999. A separate stress
evaluation draws on the calibration and tuning pool: the mixed sample comprises 1,792 feeds
associated with 256 tie lines, including 256 feeds on the liquid side of the bubble point.
A bracketing solve enforces the aqueous salt balance and the active-phase pressures. The
maximum permitted component-fraction and pressure residuals are $10^{-9}$ and 0.01~Pa,
respectively. Supplementary Table~\ref{tab:SInonPRbrine} summarizes the results for each sample.

\begin{table}[!htbp]\centering\small
\caption{\textbf{Hydrogen--brine equilibrium.} Phase-fraction and liquid hydrogen mole-fraction
MAEs use reference two-phase feeds. The maximum pressure residual uses all
feeds.}\label{tab:SInonPRbrine}
\begin{tabular}{@{}lrrrrrr@{}}\toprule
Sample & Tie lines & Feeds & Matches & $\beta$ MAE & $x_{\rm H_2}$ MAE & Max. (Pa)\\
\midrule
Held-out tie-line test & 512 & 3584 & 3584 & 0.000112 & $3.82\times10^{-5}$ & 0.000404\\
Confirmation & 512 & 3584 & 3584 & 0.0001 & $3.37\times10^{-5}$ & 0.000418\\
Mixed-feed stress test & 256 & 1792 & 1792 & $7.51\times10^{-5}$ & $4.5\times10^{-5}$ & 0.000329\\
\bottomrule\end{tabular}\end{table}

Along a fixed tie line, the feed hydrogen mole fraction varies with the vapour fraction.
Its difference from the liquid-endpoint mole fraction is $\beta(y_{\rm H_2}-x_{\rm H_2})$.
Endpoint and vapour-fraction errors therefore provide complementary measures of the accuracy
of the dilute-phase approximation.

\subsection*{5.2 Reuse of a fixed thermodynamic support set under an altered injection schedule}

The second schedule uses the same six-cell grid, thermodynamic supports, boundary pressures
and three feed compositions. The first two stages last 8,000 and 22,000~s, respectively,
and vapour-rich injection continues for the final 30,000~s. Both solvers use the same 93
time steps. Pressure, saturation and component errors are computed by pairing the two
solutions at the same step and cell. The solutions agree on all 558 phase labels and all
24 phase events, including the direction of each transition. Supplementary
Table~\ref{tab:SIflowresults} gives the two schedules side by side.

\FloatBarrier

\section*{Supplementary Note 6:\\Computing-time comparison for a six-component
equilibrium problem}

\subsection*{6.1 Equilibrium problem}

We compare the online evaluation time of MAGE with that of a compiled Peng--Robinson (PR)
solver for an equilibrium problem independent of the main-text benchmarks. The
mixture CH$_4$ + C$_2$H$_6$ + C$_3$H$_8$ + $n$-C$_4$H$_{10}$ + $n$-C$_5$H$_{12}$ + $n$-C$_6$H$_{14}$
at 300~K is described in both methods by the same PR equation of state\cite{peng1976}, with
$a_{ij}=(a_ia_j)^{1/2}$, a linear mixing rule for the covolume and zero binary interaction
parameters. The component constants are listed in Supplementary Table~\ref{tab:SIsixConstants}.

Each query specifies the temperature and the six chemical coordinates
\begin{equation}
\eta_i=\ln\frac{f_i}{p^\circ},\qquad p^\circ=10^{5}~\mathrm{Pa},\qquad i=1,\dots,6,
\label{eq:SIsixeta}
\end{equation}
where $f_i$ is the fugacity of component $i$. At fixed temperature, $\eta_i$ differs from
$\mu_i/(RT)$ only by a component-specific constant, so that $\Phi=p/(RT)$ and
$\nabla_{\boldsymbol\eta}\Phi=\boldsymbol c$ retain their meaning as the branch potential and the
vector of component molar densities. For each query, both methods return pressure, the six
component molar densities and the compressibility factor for both the liquid and vapour
branches, together with the stable phase. At prescribed temperature and chemical potentials,
the stable phase is the branch with the larger pressure because the grand potential equals
$-pV$. The PR method therefore
solves two six-dimensional nonlinear inverse problems per query, one on each root branch of the
cubic equation, whereas MAGE evaluates the two stored branch potentials and their gradients
directly. The comparison therefore characterizes chemical-potential-specified equilibrium
queries, which differ in specification from a conventional isothermal--isobaric flash at
prescribed overall composition.

\begin{table}[t]
\centering
\caption{\textbf{Component constants of the six-component Peng--Robinson mixture.}}
\small
\begin{tabular}{@{}lrrr@{}}
\toprule
Component & $T_c$ (K) & $p_c$ (MPa) & $\omega$\\
\midrule
CH$_4$ & 190.564 & 4.5992 & 0.01142\\
C$_2$H$_6$ & 305.322 & 4.872 & 0.0995\\
C$_3$H$_8$ & 369.83 & 4.248 & 0.1523\\
$n$-C$_4$H$_{10}$ & 425.125 & 3.796 & 0.2002\\
$n$-C$_5$H$_{12}$ & 469.7 & 3.370 & 0.2515\\
$n$-C$_6$H$_{14}$ & 507.6 & 3.025 & 0.3013\\
\bottomrule
\end{tabular}
\label{tab:SIsixConstants}
\end{table}

\subsection*{6.2 State sampling and data partitions}

Coexistence anchors are obtained from isothermal--isobaric PR flash calculations of the overall
composition $\boldsymbol z=(0.46,0.12,0.08,0.12,0.11,0.11)$ at 101 pressures between 1.86 and
2.84~MPa. A state is generated by drawing an anchor pressure uniformly between 1.95 and 2.75~MPa,
interpolating the corresponding liquid composition on the anchor tie lines, evaluating its chemical
coordinates and shifting the $n$-hexane coordinate by $\pm\delta$, with $\delta$ drawn uniformly
between 0.012 and 0.030 and a random sign. Each resulting chemical-potential state is inverted on
both the liquid and the vapour PR branches to a maximum residual below $10^{-8}$ in the chemical
coordinates. The states therefore form a two-parameter manifold, parameterized by the
anchor pressure
and the $n$-hexane shift, in the six-dimensional space of chemical coordinates. Sampling
on both sides of coexistence yields stable and metastable homogeneous states on each branch.

The PR samples comprise 1,200 candidate, 360 validation and 1,000 confirmatory states,
with a separate 800-state comparison pool. Of the 1,000 confirmatory states, the stable
branch is liquid for 499 and vapour for 501. The liquid and vapour compressibility factors
span 0.0733--0.1161 and 0.8873--0.9120, respectively. The smallest absolute difference
between the branch pressures is 0.0693~MPa.

\subsection*{6.3 Finite-support representation}

Each query supplies the same chemical coordinates to both branches. Preselection yields
71 native candidate planes per branch. The common rule of Supplementary Note~3 selects 60
per branch using the pressure ordering and margin of the two phases on separate calibration
and tuning states. Both branches use $\tau=0.031968$~mol\,m$^{-3}$ and share a zero offset.

\subsection*{6.4 Compiled PR solver}

The PR solver is a C++17 implementation compiled with optimization level \texttt{-O3}. For each
branch, a damped Newton method solves the six equations $\eta_i^{\mathrm{PR}}(p,\boldsymbol
x)=\eta_i$ for the logarithm of the pressure and five logarithmic composition ratios. The cubic
equation is solved analytically, and the smallest or largest admissible compressibility root is
taken for the liquid or vapour branch, respectively. Each Newton step is limited componentwise
and followed by a backtracking line search allowing up to 18 step halvings. Convergence
requires a maximum residual below $5\times10^{-11}$ in the chemical coordinates, within
at most 30 iterations. Four
variants are obtained by combining two Jacobians---an analytic Jacobian, obtained by differentiating
the PR fugacity coefficients together with the cubic root, or a forward finite-difference
Jacobian---with two initialization strategies: independent initialization of every query from a
fixed state near the centre of the sampled domain, or continuation from the converged solution of
the preceding query. On the 1,000 confirmatory queries the compiled PR pressures and
component densities agree with independent PR evaluations to maximum relative deviations
of $5.04\times10^{-10}$ and $9.31\times10^{-11}$, respectively.

\subsection*{6.5 Timing protocol and results}

Both solvers receive the same 1,000 confirmatory queries in a single batched call and return the
same output quantities. After seven warm-up passes, each timing repeat comprises 30
consecutive batched calls for every solver, with the solver order reversed between successive
repeats.
Five repeats were performed on a single process under 64-bit Windows with Clang 18.1.6 and
optimization level \texttt{-O3}. The elapsed wall time for a repeat is divided by 30,000.
All five comparisons use identical inputs and requested outputs. Supplementary
Table~\ref{tab:SIsixTiming} lists the resulting per-state times.

\begin{table}[!htbp]\centering\small
\caption{\textbf{Six-component chemical-potential queries.} Online time in $\mu$s per
state. Each repeat contains 30 batched calls on 1,000 states.}\label{tab:SIsixTiming}
\begin{tabular}{@{}lrrr@{}}\toprule
Method & Median & Min. repeat & Max. repeat\\
\midrule
MAGE & 0.5168 & 0.5124 & 0.6144\\
Analytic PR, independent & 5.739 & 5.575 & 5.985\\
Analytic PR, continuation & 5.88 & 5.873 & 5.972\\
Finite-difference PR, independent & 13.31 & 13.05 & 13.81\\
Finite-difference PR, continuation & 14.26 & 14.24 & 14.49\\
\bottomrule\end{tabular}\end{table}

\subsection*{6.6 Accuracy and scope}
All four PR implementations and MAGE assigned the correct stable branch to all 1,000 states.
Against the independent PR reference, the mean and maximum relative pressure errors of the
compiled MAGE branches were $1.65\times10^{-4}$ and $5.86\times10^{-4}$. For component
molar densities the corresponding values were 0.00378 and 0.0550. The maximum absolute
compressibility-factor error was 0.00790. The analytic PR inversion with continuation
required a median of seven total Newton steps and 16 EOS evaluations across both branches
per query. Direct MAGE evaluation took 0.517~\si{\micro\second} per query and the fastest
PR variant took 5.739~\si{\micro\second}. The measured time ratio of 11.1 thus characterizes
this two-parameter family of chemical-potential queries for the compiled implementations
described above.

\FloatBarrier

\section*{Supplementary Note 7:\\Ambient-dimension control for the support count}

In the error bounds of Supplementary Note~2, the number of supports enters through the fill distance
$\delta_M$ over the represented domain $\Omega$, together with the curvature bound $\Lambda$ and the
smoothing contribution $\bar\tau\log M$. For a regular bounded $d$-dimensional domain with
a covering-number bound of order $(D/\delta)^d$, where $D$ is a characteristic extent, the
support coverage is governed by intrinsic geometry. An isometric embedding preserves the
fill distance exactly. This control separates the ambient dimension from the intrinsic
geometry of the represented thermodynamic manifold. Liquid- and vapour-branch states of
the binary CH$_4$ +
$n$-butane system at 300~K are computed from the PR equation of state with zero binary interaction
parameter, as in the binary benchmark of the main text, and the liquid and vapour branches are
represented by the smallest and largest admissible compressibility roots, respectively. Liquid
states are sampled uniformly over pressures of 1.2--3.6~MPa and methane mole fractions of
0.055--0.22, and vapour states over 1.8--3.2~MPa and methane mole fractions of 0.84--0.94; for each
branch, 460 candidate, 280 validation and 400 test states are generated independently.
Each state supplies the chemical coordinates of Eq.~\eqref{eq:SIsixeta} for the two
components, together with the potential $\Phi=p/(RT)$ and the gradient $\boldsymbol q=(c_1,c_2)$.

The two chemical coordinates and their dual gradients are embedded in spaces of dimension
$d=2,3,5,8,10$ by $\tilde{\boldsymbol\eta}=Q\boldsymbol\eta$ and $\tilde{\boldsymbol
q}=Q\boldsymbol q$, where $Q^\mathsf T Q=I$. The potential and the value of every support
plane are invariant under this transformation because $(Q\boldsymbol q)^{\mathsf
T}Q(\boldsymbol\eta-\boldsymbol\eta_m)=\boldsymbol q^{\mathsf
T}(\boldsymbol\eta-\boldsymbol\eta_m)$. The numerical control verifies this invariance for
the selected representation. For each dimension, the common selector acts in the intrinsic
coordinates, using the pullback by $Q^\mathsf T$ on the embedded library. For the 400 test
states per phase, we project the ambient gradients back into the intrinsic coordinates
before computing their relative errors.

\begin{table}[!htbp]\centering\small
\caption{\textbf{Isometric embedding of the binary branches.} Each branch uses 83 supports
and 400 test states in each dimension. Potential errors are absolute MAEs in mol\,m$^{-3}$.
Gradient errors are mean relative vector errors.}\label{tab:SIembedding}
\begin{tabular}{@{}lrrrr@{}}\toprule
Dimension & Liquid $\Phi$ & Vapour $\Phi$ & Liquid gradient & Vapour gradient\\
\midrule
2 & 15.074 & 4.3904 & 0.019309 & 0.030937\\
3 & 15.074 & 4.3904 & 0.019309 & 0.030937\\
5 & 15.074 & 4.3904 & 0.019309 & 0.030937\\
8 & 15.074 & 4.3904 & 0.019309 & 0.030937\\
10 & 15.074 & 4.3904 & 0.019309 & 0.030937\\
\bottomrule\end{tabular}\end{table}

All five embeddings yielded identical support indices and branch errors to floating-point
precision (Supplementary Table~\ref{tab:SIembedding}). This control isolates the effect
of ambient dimension at fixed intrinsic geometry: for these isometric embeddings, the
selected supports and branch errors are invariant to the change in ambient dimension.

\FloatBarrier

\section*{Supplementary Note 8:\\Removal of supports from the selected liquid--liquid sets}

Each branch of the selected symmetric model contains 276 planes. On each branch, we selected
the three supports with the smallest maximum weights on additional homogeneous and coexistence
probes as trial candidates. We tested each single deletion and every pair consisting of
one deletion from each branch, giving 15 proposals in total. The acceptance criteria used
the mean and maximum errors in endpoint composition, total density, internal-energy density,
L2 mole fraction and L2 volume fraction. These were evaluated on nine pressures from 4.6
to 5.4~MPa and eight interleaved pressures from 4.65 to 5.35~MPa. The roundoff allowance
on each metric was $2\times10^{-14}$.

All 15 proposals exceeded at least one threshold. The retained model therefore contained
276 supports per branch and satisfied equilibrium closure at all 16 additional pressures
from 4.6125 to 5.3625~MPa. This experiment tests local redundancy under simultaneous
thermodynamic and phase-equilibrium criteria: removing any of the tested low-weight supports
increases at least one thermodynamic or equilibrium error beyond the roundoff allowance.

\FloatBarrier